\documentclass[11pt,english,a4paper,floatfix]{article}
\usepackage{jcappub}

\usepackage{bm}
\usepackage{amsfonts}
\usepackage{subfigure}

\newcommand{\be}{\begin{equation}}
	\newcommand{\ee}{\end{equation}}
\newcommand{\bea}{\begin{eqnarray}}
	\newcommand{\eea}{\end{eqnarray}}

\usepackage{lmodern}
\usepackage{slantsc}

\newcommand{\appropto}{\mathrel{\vcenter{
			\offinterlineskip\halign{\hfil$##$\cr
				\propto\cr\noalign{\kern2pt}\sim\cr\noalign{\kern-2pt}}}}}

\usepackage[scr=boondox]{mathalfa}

\usepackage{mathtools}

\usepackage{relsize,exscale}
\usepackage{array}
\usepackage{ctable}
\usepackage{cellspace} % for "\extrarowheight" macro
\newcolumntype{?}{!{\vrule width 0.12em}}

\newcommand{\CC}{C\nolinebreak\hspace{-.05em}\raisebox{.4ex}{\tiny\bf +}\nolinebreak\hspace{-.10em}\raisebox{.4ex}{\tiny\bf +}}
\newcommand{\CLASSpp}{\textsc{class}\nolinebreak\hspace{-.05em}\raisebox{.4ex}{\tiny\bf +}\nolinebreak\hspace{-.10em}\raisebox{.4ex}{\tiny\bf +}}

\usepackage{physics}

\renewcommand{\d}{\text{d}}

\begin{document}
	
%%%%%%%%%%%%%%%%%%%%%%%%%%%%%%%%%%%%%%%%%%%%%%%%%%%%%%%%%%%%%%%%%%%%%%
% Frontpage %%%%%%%%%%%%%%%%%%%%%%%%%%%%%%%%%%%%%%%%%%%%%%%%%%%%%%%%%%
%%%%%%%%%%%%%%%%%%%%%%%%%%%%%%%%%%%%%%%%%%%%%%%%%%%%%%%%%%%%%%%%%%%%%%
\title{Decaying warm dark matter revisited}

\author{Emil Brinch Holm$^a$, Thomas Tram$^a$, Steen Hannestad$^a$}

\affiliation[a]{Department of Physics and Astronomy, Aarhus University,
	DK-8000 Aarhus C, Denmark}

\emailAdd{ebholm@phys.au.dk}
\emailAdd{thomas.tram@phys.au.dk}
\emailAdd{sth@phys.au.dk}

\abstract{
	Decaying dark matter models provide a physically motivated way of channeling energy between the matter and radiation sectors. In principle, this could affect the predicted value of the Hubble constant in such a way as to accommodate the discrepancies between CMB inferences and local measurements of the same. Here, we revisit the model of warm dark matter decaying non-relativistically to invisible radiation. In particular, we rederive the background and perturbation equations starting from a decaying neutrino model and describe a new, computationally efficient method of computing the decay product perturbations up to large multipoles. We conduct MCMC analyses to constrain all three model parameters, for the first time including the mass of the decaying species, and assess the ability of the model to alleviate the Hubble and $\sigma_8$ tensions, the latter being the discrepancy between the CMB and weak gravitational lensing constraints on the amplitude of matter fluctuations on an $8 h^{-1}$ Mpc$^{-1}$ scale. We find that the model reduces the $H_0$ tension from $\sim 4 \sigma$ to $\sim 3 \sigma$ and neither alleviates nor worsens the $S_8 \equiv \sigma_8 (\Omega_m/0.3)^{0.5}$ tension, ultimately showing only mild improvements with respect to $\Lambda$CDM. However, the values of the model-specific parameters favoured by data is found to be well within the regime of relativistic decays where inverse processes are important, rendering a conclusive evaluation of the decaying warm dark matter model open to future work.
}

\maketitle

%%%%%%%%%%%%%%%%%%%%%%%%%%%%%%%%%%%%%%%%%%%%%%%%%%%%%%%%%%%%%%%%%%%%%%%%%%%%%%%%%%%%%%%%%%%%%%%%%%
\section{Introduction}
In the recent years, decaying dark matter models have received renewed interest as proposed solutions to the $4.1\sigma$ discrepancy\footnote{The discrepancy varies with the choice of local Universe observation. While tip of the red giant branch calibration of SNIa measurements give a value closer to the Planck estimate~\cite{freedman2019}, certain data combinations can increase the tension by up to $6\sigma$~\cite{divalentino2021}.} between the value of $H_0$ as inferred from Planck CMB observations, $H_0 = 67.27 \pm 0.60$ km s$^{-1}$ Mpc$^{-1}$~\cite{planck2018}, and that from local measurements by the SH0ES collaboration, $H_0 = 73.2 \pm 1.3$ km s$^{-1}$ Mpc$^{-1}$~\cite{riess2021} as well as the $\approx 3 \sigma$ discrepancy in $S_8 \equiv \sigma_8 (\Omega_m/0.3)^{0.5}$, where $\sigma_8$ denotes amplitude of matter fluctuations at $8 h^{-1}$ Mpc$^{-1}$ scales, as inferred from Planck CMB observations, $S_8 = 0.834\pm0.016$~\cite{planck2018}, and that from a joint analysis of the weak lensing surveys KIDS1000+BOSS+2dfLenS, $S_8 = 0.766^{+0.020}_{-0.014}$~\cite{heymans2020}\footnote{The tensions have been heavily reviewed in the literature; see, for example, references~\cite{schoeneberg2021, divalentino2021, perivolaropoulos2021, knox2019}. Note also that the $S_8$ tension may still be compatible with statistical fluctuations~\cite{Nunes:2021ipq}.}. It is not yet definitively clear whether these tensions arise from systematic errors, but their robustness against different local probes suggests that new physics may be required to explain the inconsistencies. It is with this motivation that we revisit decaying dark matter models and investigate their relation to the tensions.

Decaying dark matter models can be broadly classified by the nature of the decaying particle (cold or warm/hot dark matter) and the decay products (visible, massive or massless). Decays into visible decay products are strongly constrained by CMB observations~\cite{poulin2016ii, yuksel2007, zhang2007}, and arguably the most studied model is that of decaying cold dark matter (DCDM) with invisible radiation as decay products. Although studies of the latter originated several decades ago (e.g.~\cite{scherrer1985, scherrer1988i, scherrer1988ii}), analyses including the full solutions of the perturbed Boltzmann equations first arrived in the last ten years with references~\cite{audren2014},~\cite{poulin2016} and~\cite{nygaard2020, Alvi:2022aam}, who used the 2013, 2015 and 2018 data releases of Planck, respectively, to constrain the parameters of a DCDM model with invisible radiation (\emph{dark radiation}) as decay products. The strongest short-lived result, obtained by reference~\cite{nygaard2020}, shows an alleviation of the Hubble tension by about 1$\sigma$, and several other studies of cold dark matter decaying to invisible radiation agree on the ability to reduce the tension~\cite{berezhiani2015, bringmann2018, pandey2019, chen2020}. Cold dark matter decaying to massive products has also seen considerable effort, with notable progress in the works~\cite{blackadder2014, blackadder2016}, the formalism of which became convention in the later works of references~\cite{vattis2019, clark2020, haridasu2020}, until the recent references~\cite{abellan2021i, abellan2022} adopted a formalism more closely resembling that used in the massless decay product literature and concluded that decays with massive final states can alleviate the $S_8$ tension down to $1.3\sigma$\footnote{However, this is including a prior on $S_8$ from the local measurements, and noting that Bayesian model selection still favours $\Lambda$CDM.}. Currently, the strongest model parameter bounds on cold decaying model parameters are derived using the effective field theory of large scale structure~\cite{Simon:2022ftd}.

Decaying dark matter models are often grouped into those that decay at late times (i.e. well after recombination) and at early times (i.e. prior to recombination). Reference~\cite{Anchordoqui2020} argue that early time decays cannot alleviate the Hubble tension because they reduce power in the small-scale CMB damping tail, which is well constrained by data. On the other hand,~\cite{clark2020, abellan2021i} argue that late-decaying cold dark matter fails to alleviate the Hubble tension, especially if both BAO and SNIa data is included. Furthermore, several findings agree that the tensions cannot be solved simultaneously by only altering early or late time dynamics~\cite{pandey2019, nygaard2020, abellan2021i}, and the decaying cold dark matter models must be supplemented by additional extensions if they are to satisfactorily address both tensions (e.g.~\cite{clark2022}). A non-cold decaying species evades this roadblock, at least in principle, since it introduces \emph{both} early and late time changes to $\Lambda$CDM. Indeed, the early expansion history is altered due to the species being relativistic during radiation domination at early times, in contrast to the usual decaying cold dark matter models, and the late time history is altered through the decay. 

Due to modelling similarities, the history of the analysis of decaying non-cold dark matter is naturally intertwined with that of decaying neutrinos. Studies of decaying massive neutrinos emerged several decades ago (e.g.~\cite{hannestad1998i, kaplinghat1999} and references therein) when it was realized that finite neutrino lifetimes may significantly relax the bounds on the neutrino mass sum (for the most recent constraints, see~\cite{abellan2021neutrino, chen2022, barenboim2020, chacko2020}). It is only recently, with the work of reference~\cite{blinov2020}, that decaying \emph{warm} dark matter (DWDM) has been investigated explicitly. In the latter, the DWDM was seen to alleviate the Hubble tension by a modest amount, but it is still unclear exactly how it compares to the cold decaying species or other proposed solutions.

In this work, the recently derived and fully general decaying neutrino Boltzmann hierarchy of reference~\cite{barenboim2020} will be the starting point of our analysis. In particular, we will show that in the case of massless decay products and when disregarding inverse decay processes, the decaying neutrino equations of~\cite{barenboim2020} reduce exactly to the decaying warm dark matter equations used in~\cite{blinov2020}, and, by extension, to those used in the recent work~\cite{abellan2021neutrino}. In addition, we provide an approximate analytical expression for the solution to the background equations of motion for both the decaying particle and the decay products. A main contribution is a new method of computing the integrals appearing in the decay product perturbations, which is a bottleneck in the calculations~\cite{blinov2020, abellan2021neutrino}, removing the need to truncate the collision terms at a low multipole. Furthermore, we conduct Markov chain Monte Carlo (MCMC) analyses with both the initial density, lifetime and for the first time the DWDM mass in order to determine the regions of parameter space preferred by data. Lastly, we discuss the implications for both the $H_0$ and $S_8$ tensions.

This article is structured as follows. In section~\ref{sec:1.1} we give an explanation of the effect of decaying dark matter models on $H_0$. In section~\ref{sec:2}, we introduce the decaying neutrino model of~\cite{barenboim2020} from which our equations derive, and present the final equations to be solved. In section~\ref{sec:3}, we illustrate example solutions at the background and perturbative levels and discuss the impact of the decays on the CMB and matter power spectra. In section~\ref{sec:4}, we obtain parameter constraints from MCMC analyses and discuss the impact of DWDM on the $H_0$ and $S_8$ tensions, after which we conclude on our findings in section~\ref{sec:5}.

\subsection{Why does decaying dark matter increase $H_0$ relative to $\Lambda$CDM?}\label{sec:1.1}
\begin{figure}[tb]
	\centering 
	\includegraphics[width=\textwidth]{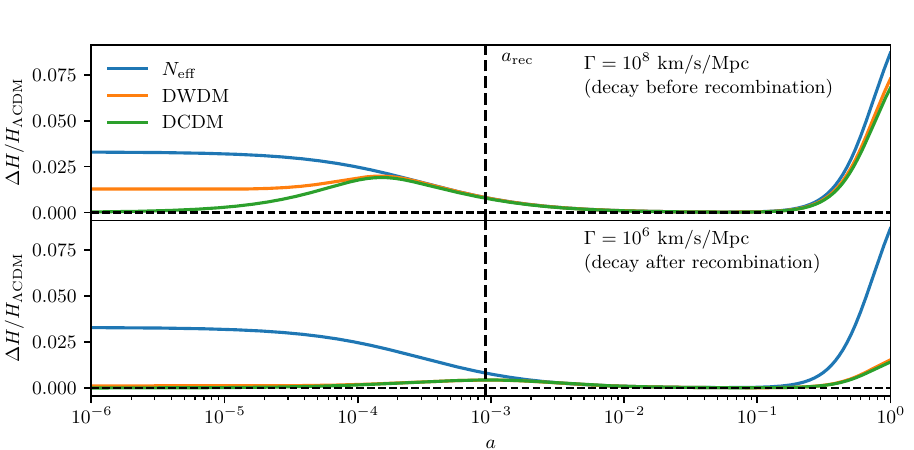}
	\caption{\label{fig1}Relative change in the Hubble parameter $H$ as computed by \textsc{class}~\cite{CLASS2} for several models, as a function of scale factor $a$. The red, yellow and green lines represent a warm decaying species with mass $m=10$ eV, a dark radiation species with radiation density corresponding to $\Delta N_{\text{eff}}=0.5$ today and a cold decaying species, respectively. The decaying models use the same decay rate of $\Gamma =10^8$ km s$^{-1}$ Mpc$^{-1}$ and $\Gamma =10^6$ km s$^{-1}$ Mpc$^{-1}$ for the top and bottom panels, respectively, and have their densities fixed such that their decay products contribute additional radiation corresponding to $\Delta N_{\text{eff}}=0.5$ today. The acoustic scale is fixed at $\theta_s = 1.042143$, and the value of $H_0$ is inferred from self-consistency.}
\end{figure}

Decaying dark matter models, be they cold or warm, are all seen to increase the Hubble constant $H_0$ relative to its $\Lambda$CDM value. This result is usually seen directly from a Markov chain Monte Carlo parameter inference. In this section, we detail the physical intuition behind this increase. The effect is illustrated on figure~\ref{fig1}. Here, the fractional change in the Hubble parameter is given for a DWDM model with decaying particle mass $m=10$ eV along with a DCDM model and a purely massless dark matter species parametrized by a shift in the effective number of neutrino species $\Delta N_{\text{eff}}$. All models have been matched such that they contribute an additional radiation density \emph{today} corresponding to $\Delta N_{\text{eff}} = 0.5$ (consequently, they have the same cosmological constant) and the warm and cold decaying models use the same decay rate of $\Gamma=10^8$ km s$^{-1}$ Mpc$^{-1}$ and $\Gamma=10^6$ km s$^{-1}$ Mpc$^{-1}$ for the top and bottom panels, corresponding to decays occurring before and after recombination, respectively. Also indicated on the figure is the scale factor $a_{\text{rec}}$ of recombination. Furthermore, the acoustic scale is fixed near its observational value $\theta_s = 1.042143$~\cite{planck2018} in order to allow $H_0$ to vary; we find this to be a natural choice since $\theta_s$ is highly constrained, model independently, by Planck data~\cite{planck2018}.

As is seen on the top panel with pre-recombination decays, at early times, the Hubble parameters of the DWDM and dark radiation models are larger than the $\Lambda$CDM and DCDM models, since the former introduce additional relativistic energy density, which dominates in the early Universe (note that the parameters used in the figure are such that the DWDM species becomes non-relativistic at around $a\approx 10^{-4}$, hence its relativistic energy contribution). Contrarily, the DCDM species only contributes radiation energy through its decay products, so its early-time value of $H$ is similar to that from a $\Lambda$CDM cosmology. In the bottom panel, the early DWDM energy density is smaller because the decay happens later and we fix the final energy in the sector; as we will argue, this diminishes the late-time decrease in $H(z)$. In this sense, the early-time values of $H$ indicate that the DWDM model can be seen as an interpolation between DCDM and pure $\Delta N_{\text{eff}}$; a theme we will explore further in section~\ref{sec:3}. Once the DWDM species becomes non-relativistic, the Hubble parameters of the DWDM and DCDM models converge, after which they in turn converge to the $\Delta N_{\text{eff}}$ model after decaying into massless decay products. Furthermore, in all three models, $H$ converges to the $\Lambda$CDM value just after this time since the additional radiation redshifts into negligence during the epoch of matter domination.

The reason for the late-time increase in $H$ can be understood by an argument from references~\cite{knox2019,anc,schoeneberg2021}. The acoustic scale $\theta_s$ is related to the sound horizon at last scattering $r_s (z_*)$ and the comoving angular diameter distance $D_A(z_*)$ to recombination by
\begin{align} \nonumber
	\theta_s = \frac{r_s (z_*)}{D_A(z_*)}, \quad \text{ where } \quad r_s (z_*) = \int_{z_*}^{\infty} \frac{c_s (z')}{H(z')} \text{d}z' \quad \text{ and } \quad D_A(z_*) = \int_{0}^{z_*} \frac{1}{H(z')} \text{d}z',
\end{align}
with $c_s$ denoting the sound speed. As mentioned, the $\Delta N_{\text{eff}}$ and DWDM models have a larger value of $H(z)$ in the early, radiation dominated Universe, since they both include additional relativistic components, as does the DCDM model once the decay sets in and produces dark radiation. This increase in $H(z)$ for $z>z_*$ yields a decreased sound horizon at last scattering, as can be seen by its integral representation, since the sound speed is dependent on $\omega_b$ alone~\cite{knox2019}. Consequently, to keep $\theta_s$ fixed, the angular diameter distance to recombination $D_A (z_*)$ must increase, which implies an increased value of $H(z)$ for $z < z_*$. Since $H(z)$ falls off rapidly with decreasing $z$, the largest contribution to the integral comes from $H(z)$ at small $z$, explaining the sudden increase in $H$ for late times seen in figure~\ref{fig1}. The upshot is that all models predict a larger $\Delta H_0 \equiv H_0 - H_{0,\Lambda\text{CDM}}$ relative to the value at recombination $\Delta H(z_*)$, which illustrates their collective ability to increase the value of $H_0$ as interpolated from a measurement around the epoch of recombination.

%%%%%%%%%%%%%%%%%%%%%%%%%%%%%%%%%%%%%%%%%%%%%%%%%%%%%%%%%%%%%%%%%%%%%%%%
\section{Theory} \label{sec:2}
In this section, we show how the formalism of a DWDM species can be derived from a decaying neutrino model and present the set of Boltzmann equations governing its evolution at both background level and to first order in perturbation theory.

\subsection{Physical system}
A fundamental model enabling a decaying dark matter species is equivalent to the decaying neutrino model studied in reference~\cite{barenboim2020}, where a universal interaction between two neutrino species and a light scalar particle is introduced via the effective Lagrangian
\begin{align} \label{lagrangian}
	\mathcal{L}_{\text{int}} = \mathfrak{g}_{ij} \overline{\nu}_i \nu_j \phi,  
\end{align}
where $\mathfrak{g}_{ij} = \mathfrak{g}$ is a universal coupling constant. This interaction term permits three $2+2$ scattering processes and a single $2+1$ decay-type process,
\begin{align} \nonumber
	\underbrace{\nu \nu \leftrightarrow \nu \nu, \quad \nu \phi \leftrightarrow \nu \phi, \quad \nu \nu \leftrightarrow \phi \phi}_{\text{scattering}}, \quad \underbrace{\nu_H \leftrightarrow \nu_l + \phi}_{\text{decay}}
\end{align}
where, in the latter, the subscripts $H$ and $l$ indicate particles with heavy and light masses, respectively. The three scattering processes have been studied extensively, e.g. in references~\cite{oldengott2014, oldengott2017}. In this work, we disregard the contributions from these and focus only on the decay process. As argued in reference~\cite{barenboim2020}, the interaction rate of the scattering processes scales as $\Gamma_\text{scatter} \sim \mathfrak{g} T^4$ (focusing on a single neutrino species). On the other hand, the rest frame decay rate of the decaying particle is $\Gamma_{\text{dec}} = \mathfrak{g}^2 m_H / 4 \pi$~\cite{barenboim2020}, assuming it is a Majorana particle. Crucially, for non-cold dark matter populations, this is time-dilated by the Lorentz factor $ \gamma = E_H / m_H$, where $E_H = (m_H^2 + p^2)^{1/2}$ is the energy of a single particle with rest mass $m_H$ and momentum $p$. Given that the mean momentum of a relativistic particle in the early Universe is $p \approx 3.15 T$~\cite{kolb}, and taking this as a suggestive momentum of the population, we get a temperature dependence of the decay interaction rate on the form $\Gamma_{\text{dec}} \sim 1/(m_H^2 + T^2)^{1/2}$. Contrary to the interaction rate of the scattering processes, this rate increases with time. We therefore expect a late time epoch where the decays dominate and scattering can be neglected. Reference~\cite{barenboim2020} show that at around $\mathfrak{g} = 10^{-10}$, the decay process dominates for all temperatures below roughly $10^7$ eV, so the scattering processes may well be neglected in this small-$\mathfrak{g}$ regime.

Despite being derived from a decaying neutrino model, our analysis is agnostic about the particle physics realization of the decaying species. The model agnosticism is a powerful asset of our model, implying that the current analysis applies to many examples found in the literature, such as early hot dark sectors~\cite{Ertas:2021xeh}, sterile neutrino decays into majorons and the subsequent decays of majorons into neutrinos~\cite{Escudero:2019gvw, Escudero:2021rfi}, as well as neutrino decays realized by a range of particle physics models~\cite{escudero2020, escudero2019, chacko2020}. The only immediate restriction is in the choice of initial conditions, although it is straightforward to implement changes to these initial conditions in our code if desired. In this work, we will take a Fermi-Dirac distribution with the Standard Model bath temperature as initial condition for the decaying particle, scaled by some constant degeneracy parameter, as realizable for example through the Dodelson-Widrow mechanism~\cite{dodelson_widrow}. Alternatively, it has been shown that given a sufficiently large mixing angle and a lepton asymmetry close to zero, the eV-scale decaying sterile neutrinos may indeed thermalize with the Standard Model bath, justifying taking a Fermi-Dirac initial distribution with a temperature $T \approx T_{\text{SM}}$~\cite{hannestad2012}. As for the decay products $\phi$, we assume that the entire population stems from the decays. This is realizable in most of the mentioned models~\cite{blinov2020}, and since we restrict ourselves to the regime of small coupling strengths, any $\phi$ production through the scattering interactions will be negligible. Finally, note that the decaying species is only expected to make up a fraction $f_{\text{dwdm}} = \Omega_{\text{dwdm}}/\Omega_{\text{dm}}$ of all dark matter. Therefore, it escapes the lower bounds on its mass, typically of a few keV, as provided e.g. by Lyman-$\alpha$ studies and the Tremaine-Gunn bound~\cite{white_paper_kev}.

The phenomenology of the species depends heavily on whether it decays while relativistic or while non-relativistic. In particular, if the species decays appreciably while relativistic, the inverse decay process $\nu_l + \phi \rightarrow \nu_H$ will be energetically feasible, and once a substantial population of decay products has been established, it will commence at a similar rate to the decay~\cite{barenboim2020}. On the other hand, if $\nu_H$ is non-relativistic, the energy of its decay products will redshift faster than its own, suppressing the rate of the inverse decay. The nature of the decay may be classified through the relativity parameter~\cite{hannestad1998iii},
\begin{align} \label{relativity_parameter}
	\alpha = 3.5 \left( \frac{m_H}{\text{eV}} \right)^2 \frac{\tau}{10^6 \text{ yr}},
\end{align}
where $\tau = 1/\Gamma_{\text{dec}}$ denotes the lifetime. Indeed, since a massive species becomes non-relativistic roughly at a temperature $T_{\text{nr}} \approx m_H /3.15$, it follows that the particle largely decays while relativistic if $\alpha \lesssim 1$ and decays non-relativistically otherwise. In this work, we will only consider non-relativistic decays, and an evaluation of $\alpha$ will therefore guide our choice of prior ranges when conducting Bayesian inference in section~\ref{sec:4}. 

\subsection{Boltzmann equations}
In this section, we present the equations of motion arising from the proposed Lagrangian (\ref{lagrangian}). As usual, we expand the metric in terms of small perturbations around a homogeneous and isotropic background~\cite{ma1995},
\begin{align} \nonumber
	\d s^2 = a(\tau)^2 \left[ - \d \tau^2 + (\delta_{ij} + h_{ij}(\bm{x}, \tau) \d x^i \d x^j) \right],
\end{align}
where $\tau$ denotes conformal time and we will work in synchronous gauge for the entirety of this paper. Furthermore, we expand the distribution function $f_i (\bm{x}, P^\mu, \tau)$ (giving the number count per infinitesimal phase space volume) in terms of a homogeneous, isotropic part $\overline{f}_i (p, \tau)$ and a small inhomogeneous part $\Psi_i (\bm{x}, \bm{p}, \tau)$,
\begin{align} \nonumber
	f_i (\bm{x}, P^\mu, \tau) = \overline{f}_i (p, \tau) \left[ 1 + \Psi_i (\bm{x}, \bm{p}, \tau) \right],
\end{align}
where $P^\mu$ denotes the four-momentum, $\bm{p}$ the three-momentum and $i$ indexes some species. The fundamental equation governing the evolution of the metric and the distribution functions is the relativistic Boltzmann equation~\cite{barenboim2020},
\begin{align} \label{boltzmann}
	P^\mu \pdv{f_i}{x^\mu} - \Gamma^\nu_{\rho \sigma} P^\rho P^\sigma \pdv{f_i}{P^\nu} = \frac{\epsilon_i}{a^2} \left( \dv{f_i}{\tau} \right)_C,
\end{align}
where $\Gamma^\nu_{\rho \sigma}$ denotes the Christoffel symbols, $m_i$ the mass of the $i$'th species, $\epsilon_i \equiv a (p_i^2 + m_i^2)^{1/2}$ the comoving single particle energy and the collision term on the right hand side is derived from the interactions specified in the last section. Since the latter has been worked out by reference~\cite{barenboim2020}, we will not reiterate it here explicitly. By substituting the perturbed ansatz of the metric and distribution functions and equating terms of same order in the above, one obtains independent sets of evolution equations for the homogeneous, isotropic quantities (the \emph{background} quantities) and the first order inhomogeneous, anisotropic quantities (the \emph{perturbed} quantities). In the next subsections, we discuss these equations individually.

\subsubsection{Background equations} \label{sec:2.2.1}
As mentioned above, reference~\cite{barenboim2020} computed the collision term on the right hand side of equation (\ref{boltzmann}). In the special case of massless decay products and excluding inverse decays and quantum statistical effects, the evolution of the homogeneous and isotropic part of the decaying particle distribution function $\overline{f}_H$, given in equation (4.12) of reference~\cite{barenboim2020}, reduces to
\begin{align} \label{mother_back}
	\pdv{\overline{f}_H (q_1)}{\tau} = - \frac{a^2 m_H \Gamma}{\epsilon_1} \overline{f}_H (q_1)
\end{align}
where $q_i \equiv a|\bm{p}_i|$ and the decay rate $\Gamma$ is related to the coupling constant by $\Gamma = \mathfrak{g}^2 m_H/4\pi$. By integrating over $a^{-4} \mathrm{d}^3 \bm{q}_1 \epsilon_1$ this can be recast in terms of the evolution of the energy density as 
\begin{align} \label{mother_dens}
	\dv{\rho_H}{\tau} + 3 a H (\rho_H + p_H) = -a\Gamma m_H n_H,
\end{align}
with $p_H$ denoting the homogeneous and isotropic pressure and $n_H$ the number density of the decaying species. Note that the right hand side contains the factor $m_H n_H$, and not $\rho_H$, as usually seen in the equations of a decaying cold species~\cite{nygaard2020, poulin2016}. Of course, for small values of $q$, $\rho=mn$, so the above generalizes the DCDM equation. As we will see, tracking the energy density of the decay products requires knowing the distribution function $\overline{f}_H (q)$ at each value of the comoving momentum $q$, so we have numerically implemented equation (\ref{mother_back}) rather than the integrated equation (\ref{mother_dens}).

Taking the special case for the decay products is less trivial and involves some analytical work. We present the full derivation in appendix~\ref{apA}. In particular, we combine the two decay products $\nu_l$ and $\phi$ into a single fluid, henceforth dubbed \emph{dark radiation}, with distribution function $\overline{f}_{\text{dr}} \equiv (2 \overline{f}_l + \overline{f}_\phi)/2$ being the spin-weighted sum of decay products (the factor $1/2$ balances the fact that we have implicitly removed a spin-factor $2$ from the decaying species). The resulting equation for the background distribution function is 
\begin{align} \label{daughter_back}
	\dv{\overline{f}_{\text{dr}}(q_2)}{\tau} = \frac{2 a^2 m_H \Gamma}{q_2^2} \int_{q_{1-}}^{\infty} \mathrm{d}q_1 \frac{q_1}{\epsilon_1} \overline{f}_H(q_1), \quad q_{1-} = \left| \frac{a^2 m_H^2}{4q_2} - q_2 \right|.
\end{align}
The nontrivial lower integral bound provides some interesting physical insight: It converges to infinity in the limits $q_2 \rightarrow 0$ and $q_2 \rightarrow \infty$, and has a minimum at $q_2 = am_H/2$. That is, the integration region is largest exactly when the two decay products receive the same momentum. This can be understood from the kinematics of the decay; momentum conservation requires that each decay product be created with momentum $p=m_H/2$, which then redshifts with the inverse scale factor. Hence, a single decay at scale factor $a_D$ populates a momentum bin with $q=a_D m_H/2$, and, among other things, we expect the peak of the decay product distribution $\overline{f}_{\text{dr}} (q_2)$ to correlate with the time at which most particles decayed. Ultimately, we conclude that it is in fact the lower integral bound $q_{1-}$ that enforces momentum conservation in practice, and therefore stress the importance of a robust way to implement it numerically.

For the analysis of this paper, however, it is sufficient to track only the energy density of the dark radiation. Since the lower integral bound itself depends on the momentum $\bm{q}_2$, the integration requires some analytical work, which is again detailed in appendix~\ref{apA}. In the end, we arrive at the expected result
\begin{align} \label{daughter_dens}
	\dv{\rho_{\text{dr}}}{\tau} + 4 aH \rho_{\text{dr}} = a \Gamma m_H n_H.
\end{align}
We note that equations (\ref{mother_back}), (\ref{mother_dens}) and (\ref{daughter_dens}) agree simultaneously with the decaying dark matter equations in reference~\cite{blinov2020} and the decaying neutrino equations in reference~\cite{abellan2021neutrino}.

The factor $1/\epsilon_1$ in equation (\ref{mother_back}) means that the distribution function $\overline{f}_H$ decays at different rates for each momentum bin, which is a direct manifestation of time dilation. As a consequence, the equation has no closed analytical solution in terms of elementary functions for a given power law Universe $a(t) = \kappa t^{2/3+3w}$ for some dominant equation of state parameter $w$, with $t$ denoting cosmic time. However, since we model only a non-relativistically decaying species, one can assume a stable evolution while relativistic and then expand the equation in $q/\epsilon$ while non-relativistic. This approximation scheme, detailed further in appendix~\ref{apA2}, admits an analytically closed expression for the densities at all times, from which an approximate relation between the total energy density parameter $\Omega_{0,\text{dwdm+dr}}$ of the decaying sector today and the initial energy density
\begin{align} \label{shooting_guess}
	\Omega_{0,\text{dwdm+dr}} = \Omega_{\text{ini,dwdm}} \frac{\kappa \Gamma \exp(\Gamma t_{\text{nr}})}{a(t_{\text{nr}})} \left( t_{\text{nr}}^{\frac{5 + 3w}{3+3w}} E_{\frac{-2}{3+3w}} (\Gamma t_{\text{nr}}) -  t_0^{\frac{5 + 3w}{3+3w}} E_{\frac{-2}{3+3w}} (\Gamma t_0)\right),
\end{align}
where $E_k (x)$ denotes the generalized exponential integral of variable order $k$, $a(t_\text{nr}) \approx 3.15 T/m$ is the scale factor of the non-relativistic transition and $\Omega_{\text{ini},\text{dwdm}}\equiv \rho_{H,\text{ini}}a_{\text{ini}}^4/\rho_{\text{crit}}$ denotes the total current energy density parameter of the decaying sector \emph{if the decaying species were stable}~\cite{audren2014}. This solution omits any feedback of the DWDM species on the scale factor evolution. For realistic values of the total densities of the decaying sector, however, we find that the error induced by this is only a few percent. In practice, we find that (\ref{shooting_guess}) predicts the correct final density within a factor $\approx 4$ for the relevant regions in parameter space, which is satisfactory for use as a starting point in the shooting algorithm of \textsc{class}.

\subsubsection{Perturbation equations}
In this section, we present the equations for the distribution function inhomogeneities and isotropies to first order in perturbation theory. As usual, we decompose the Fourier transforms of the distribution function perturbations $\Psi_i (\bm{k}, \bm{q}_i, \tau)$ in terms of Legendre polynomials $P_\ell (\hat{k}\cdot \hat{q}_i)$, 
\begin{align} \nonumber
	\Psi_i (\bm{k}, \bm{q}_i, \tau) = \sum_{\ell  = 0}^{\infty} (-i)^\ell (2\ell + 1)\Psi_{i, \ell} (k, q_i) P_\ell (\hat{k} \cdot \hat{q}_i),
\end{align}
and obtain an infinite sequence of equations for the multipole moments $\Psi_{i, \ell}$,
\begin{align}
	\dot{\Psi}_{i, 0} (q_i) &= - \frac{q_i k}{\epsilon_i} \Psi_{i, 1} (q_i) + \frac{\dot{h}}{6} \pdv{\ln \overline{f}_i (q_i)}{\ln q_i} + \mathcal{C}^{(1)}_0 [\Psi_i (q_i)] \nonumber \\
	\dot{\Psi}_{i, 1} (q_i) &= \frac{q_i k}{\epsilon_i} \left( -  \frac{2}{3} \Psi_{i, 2} (q_i) + \frac{1}{3} \Psi_{i, 0} (q_i) \right) + \mathcal{C}^{(1)}_1 [\Psi_i (q_i)] \label{mother_hierarchy} \\
	\dot{\Psi}_{i, 2} (q_i) &= \frac{q_i k}{\epsilon_i} \left( -  \frac{3}{5} \Psi_{i, 3} (q_i) + \frac{2}{5} \Psi_{i, 1} (q_i) \right) - \pdv{\ln \overline{f}_i (q_i)}{\ln q_i} \left( \frac{2}{5} \dot{\eta} + \frac{1}{15} \dot{h} \right) +  \mathcal{C}^{(1)}_2 [\Psi_i (q_i)] \nonumber \\
	\dot{\Psi}_{i, \ell} (q_i) &= \frac{k}{2\ell + 1} \frac{q_i}{\epsilon_i} \left[ \ell \Psi_{i, \ell-1} (q_i) - (\ell + 1) \Psi_{i, \ell+1}(q_i) \right] + \mathcal{C}^{(1)}_\ell [\Psi_i (q_i)], \qquad \ell \geq 3, \nonumber
\end{align}
where dots denote derivatives with respect to $\tau$, $h \equiv h^i_i (\bm{k}, \tau)$ is the trace of the Fourier transformed metric perturbation and, due to the possibility of a dynamical background distribution, the effective collision terms $\mathcal{C}^{(1)}_\ell [\Psi_i (q_i)]$ contain two terms,
\begin{align} \nonumber
	 \mathcal{C}^{(1)}_\ell [\Psi_i (q_i)] \equiv \frac{1}{\overline{f}_i } \left( \dv{f_i}{\tau} \right)_{C, \ell}^{(1)} - \frac{1}{\overline{f}_i} \dv{\overline{f}_i}{\tau} \Psi_{i, \ell}.
\end{align}
In the case of the decaying particle, \emph{the two terms in the above exactly cancel}~\cite{barenboim2020}, so there are no collision terms when disregarding inverse processes. This is a direct manifestation of the fact that the decay is a background process.

In the case of the decay products, one can use the fact that they are massless to average the Boltzmann equation over $\d q \ q^2 q \overline{f}_\text{dr} (q)$ so that one escapes explicit calculations of the momentum dependent perturbations $\Psi_\text{dr} (q)$ in favour of the momentum averaged quantities~\cite{ma1995}, 
\begin{align} \label{mom_avg}
	F_{\text{dr}} (\bm{k}, \tau) \equiv r_{\text{dr}} \frac{\int q^2 \d q \ q \overline{f}_{\text{dr}}(q, \tau)\Psi_{\text{dr}} (\bm{k}, \bm{q}, \tau) }{\int q^2 \d q \ q \overline{f}_{\text{dr}} (q, \tau)}.
\end{align}
After decomposing these in terms of Legendre polynomials, the resulting infinite sequence of equations for the multipole moments $F_{\text{dr}, \ell}$ is
\begin{align} \nonumber
	\dot{F}_{\text{dr},0} &= -k F_{\text{dr}, 1} - \frac{2}{3} r_{\text{dr}} \dot{h} + \left( \dv{F_\text{dr}}{\tau} \right)_{C, 0}^{(1)} \\
	\dot{F}_{\text{dr},1} &= \frac{k}{3} F_{\text{dr},0} - \frac{2k}{3} F_{\text{dr},2} +  \left( \dv{F_\text{dr}}{\tau} \right)_{C, 1}^{(1)}  \label{massless_hierarchy} \\
	\dot{F}_{\text{dr},2} &= \frac{2k}{5}  F_{\text{dr},1} - \frac{3k}{5}  F_{\text{dr},3} + \frac{4}{15} r_{\text{dr}} \left( \dot{h} + 6\dot{\eta}\right) +  \left( \dv{F_\text{dr}}{\tau} \right)_{C, 2}^{(1)} \nonumber \\
	\dot{F}_{\text{dr},\ell} &= \frac{k}{2\ell + 1} \left( l F_{\text{dr},\ell-1} - (\ell+1) F_{\text{dr},\ell+1}  \right) +  \left( \dv{F_\text{dr}}{\tau} \right)_{C, \ell}^{(1)}, \quad \ell \geq 3   \nonumber .
\end{align}
The advantage of the momentum averaging is of course that one does not need to track individual momentum bins for each multipole moment of the perturbations, but only the momentum averaged perturbation at each multipole, reducing a large part of the computational cost of solving the equations numerically.

Reference~\cite{barenboim2020} worked out the perturbation collision terms for the decay products in details. In appendix~\ref{apB}, we carry out the momentum integration to obtain the momentum averaged collision terms appearing in the hierarchy above. The result is
\begin{align} \label{col_int}
	 \left( \dv{F_\text{dr}}{\tau} \right)_{C, \ell}^{(1)} = \dot{r}_{\text{dr}} \frac{\int_0^\infty \d q \ q^2 \overline{f}_H(q) \Psi_H (q) \mathcal{F}_\ell (q/\epsilon)}{\int_0^\infty \d q \ q^2 \overline{f}_H(q)},
\end{align}
where $\dot{r}_{\text{dr}} \equiv \d (\rho_{\text{dr}} a^4 / \rho_{\text{crit}})/\d \tau = r_{\text{dr}} \ a \Gamma m_H n_H / \rho_{\text{dr}}$. A similar result was obtained in reference~\cite{blinov2020}. Here, the scattering kernel $\mathcal{F}_\ell (x)$ is the integral
\begin{align} \nonumber
	\mathcal{F}_\ell (x) \equiv \frac{(1 - x^2)^2}{2} \int_{-1}^{+1} \frac{P_\ell (u)}{(1 - xu)^3} \d u. 
\end{align}
The momentum integral (\ref{col_int}) over the decaying particle distribution is expensive to evaluate numerically, particularly so if one evaluates the kernel $\mathcal{F}_\ell (x)$ by explicit integration, as has been done in previous works. However, the integral in $\mathcal{F}_\ell$ may be computed analytically using equation (7.228) of reference~\cite{gradshteyn},
\begin{align} \nonumber
	\int_{-1}^{+1} \frac{P_\ell (u)}{(z - u)^{\mu + 1}} \d u = \frac{2 (z^2 - 1)^{-\mu/2}}{\Gamma(1 + \mu)} e^{-i\pi \mu} Q^\mu_\ell (z),
\end{align}
where $\Gamma$ denotes the Gamma function and $Q^\mu_\ell$ denotes the associated Legendre polynomial of the second kind. Using this, substituting $z = x^{-1}$ and setting $\mu=2$, we find 
\begin{align} \nonumber
	\mathcal{F}_\ell (x) = \frac{1 - x^2}{2x} Q^2_\ell \left(\frac{1}{x}\right).
\end{align}
Ordinarily, with $z$ real, $Q^2_\ell (z)$ is only defined for $ |z| < 1$. In our case, $z = 1/x$ is evaluated at $x=q/\epsilon$, where $q, \epsilon$ denote comoving momentum and energy, respectively. Therefore, we will generally have $z > 1$, and must extend the domain of $Q^2_\ell$ by analytic continuation, choosing a branch without branch cuts on the positive real axis. One such exists; it has the branch cut $(-\infty, -1)$, well outside the domain of interest.

$Q^2_\ell$ can be written in terms of the hypergeometric function and inherits a useful recurrence relation from it~\cite{abramowitz},
\begin{align} \nonumber
	(\ell - 2) Q_\ell^2 (z) = (2\ell - 1) z Q_{\ell - 1}^2 (z) - (\ell + 1) Q_{\ell - 2}^2 (z)
\end{align}
which the scattering kernel in turn inherits directly,
\begin{align} \label{F_recurrence}
	(\ell - 2) \mathcal{F}_\ell (x) = (2\ell - 1) \frac{1}{x} \mathcal{F}_{\ell - 1} (x) - (\ell + 1) \mathcal{F}_{\ell - 2} (x).
\end{align}
It is valid for $\ell \geq 3$, since the left hand side vanishes with $\ell = 2$. The values for $\ell = 0, 1, 2$ are therefore computed individually; written out, they are
\begin{align} \nonumber
	\mathcal{F}_0 (x) = 1, \quad \mathcal{F}_1 (x) = x, \quad \mathcal{F}_2 (x) = \frac{x (5x^2 -3) + 3(x^2 - 1)^2 \tanh^{-1} (x)}{2x^3}.
\end{align}
As it turns out, forwards recurrence is unstable for almost all values of interest so in these cases we switch to backwards recurrence using Miller's algorithm~\cite{Press2007}. This provides an efficient method of computing $\mathcal{F}_\ell (x)$ with potentially large $l_{\text{max}}$, since the amount of terms in $\mathcal{F}_\ell (x)$ increases rapidly with the multipole.

%%%%%%%%%%%%%%%%%%%%%%%%%%%%%%%%%%%%%%%%%%%%%%%%%%%%%%%%%%%%%%%%%%%%%%%%
\section{Numerical solutions} \label{sec:3} 
We have implemented the equations described in the last section in the code \CLASSpp{}, a translation of the Einstein-Boltzmann \textsc{class} to the programming language \CC{}. In this section, we describe our implementation, present sample solutions and discuss the impact of the model on observables. Unless otherwise is stated explicitly, the figures in this section are produced with a background $\Lambda$CDM cosmology, a decay constant $\Gamma = 10^8$ km s$^{-1}$ Mpc$^{-1}$ and an energy density scaled such that the contribution from the decay products to $N_{\text{eff}}$ today is $0.5$.

\subsection{Background equations} \label{sec:3.1}
In this subsection, we detail the numerical implementation of the background equations. The code evolves the decaying species as a non-cold dark matter species and takes as input a decay constant $\Gamma$ (or lifetime $\tau = \Gamma^{-1}$), its mass $m$ and one of several parameters measuring the energy density of the decaying sector; in this work, we will mainly use the contribution of the decaying species to the effective number of neutrino species at initial time, $N_\text{eff,ini}$, because it has a somewhat straightforward interpretation. Since the decaying species is relativistic and assumed to have decayed a negligible amount at initial time, $N_\text{eff,ini}$ is directly related to the initial energy density. In order to close the Friedmann equation, both the initial and final energy density must be known: Our code finds one from the other self-consistently using a shooting algorithm.

In order to trace the exponentially decaying distribution function $\overline{f}_H (q)$ beyond the point at which it becomes smaller than machine precision, we instead evolve its natural logarithm, whose magnitude is always well within machine precision. This allows the computation of ratios of exponentially decaying quantities even long after the distribution function $\overline{f}_H (q)$ is below machine precision by expanding the ratio with a suitable constant before exponentiating $\ln \overline{f}_H$. Important examples are the equation of state $w$ and moments of the perturbed distribution function such as $\delta$ and $\theta$.

Contrary to the case of stable non-cold dark matter species (and decaying cold dark matter, as we will see below) the shape of the distribution function also changes with time. Since the sourcing of the perturbations depends on this (\ref{mother_hierarchy}), we also record it across the momentum grid at each point in conformal time. The logarithmic derivative of $\ln \overline{f}_H(q)$ is computed by interpolation; this is particularly straightforward since we solve directly for $\ln \overline{f}_H(q)$. To combat potential stiffness of this system of equations, we employ the stiff integrator \texttt{ndf15} in \textsc{class} also for the background computations. 

\begin{figure}[tb]
	\centering 
	\includegraphics[width=\textwidth]{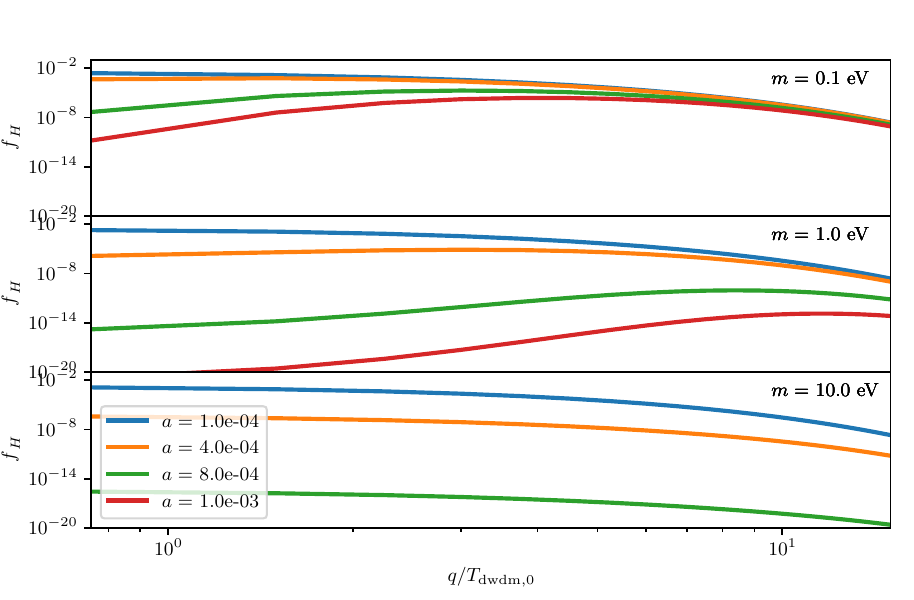}
	\caption{\label{fig3} DWDM distribution functions for $\Gamma = 10^8$ km s$^{-1}$ Mpc$^{-1}$ and three masses at different scale factor values. It is seen that the $m=0.1$ eV and $m=1.0$ eV species display non-uniform decay; the distribution function at large momenta survives longer due to time dilation of the lifetime. On the other hand, the $m=10.0$ eV species approaches the DCDM limit and therefore decays roughly uniformly across the momenta. Note that the momenta on the first axis are normalized by the temperature of the species today, $T_{\text{dwdm},0}$, taken to be equal to the neutrino temperature today.}
\end{figure}

Figure~\ref{fig3} shows the decaying particle distribution function $\overline{f}_H (q)$ at different scale factors around the decay time for three masses $m=0.1$ eV$, 1.0 $ eV and $10.0$ eV. For the masses $m=0.1$ eV and $m=1.0$ eV, it is clearly seen that the distribution function at large momenta decays slower than at small momenta, which was also predicted from the momentum dependent denominator in equation (\ref{mother_back}). This non-uniform decay is a defining characteristic of decaying warm dark matter as opposed to decaying cold dark matter. Indeed, for the latter, the single particle energy is made up only of the mass, $\epsilon = m_H a$, and equation (\ref{mother_back}) reduces to $\dot{\overline{f}}_H (q) = -am_H \Gamma \overline{f}_H (q)$, such that each momentum bin has the same exponential decay constant. Physically, the delayed decay of the particles with large momenta is of course a manifestation of the time dilation of their lifetime. However, with increasing mass, the warm particles converge to the decaying cold dark matter limit. As seen on figure~\ref{fig3}, this already starts to set in at $m=10.0$ eV, for which the decay occurs almost uniformly across the momentum bins. 

\begin{figure}[tb]
	\centering 
	\includegraphics[width=\textwidth]{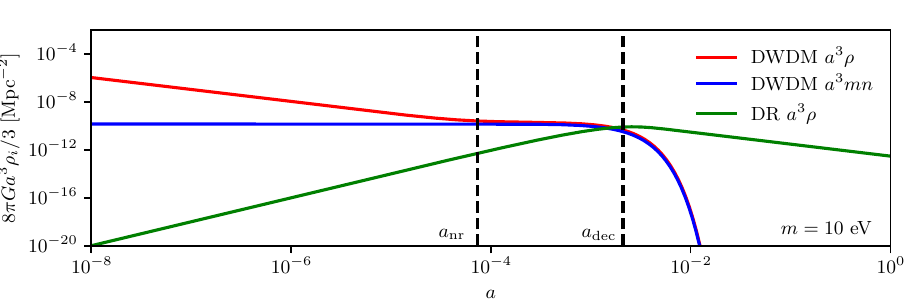}
	\caption{\label{fig4} Evolution of the background densities for a DWDM species with $m=10.0$ eV and $\Gamma = 10^6$ km s$^{-1}$ Mpc$^{-1}$. The DWDM energy density converges to the rest mass energy as the species becomes non-relativistic, and decays exponentially at a later point. }
\end{figure}

The energy density of the decaying species is obtained by straightforward integration of the distribution function. Figure~\ref{fig4} illustrates the energy density $\rho_H$ of a $\Gamma = 10^6$ km s$^{-1}$ Mpc$^{-1}$, $m=10.0$ eV DWDM species, its rest mass energy density $m_H n_H$ and the decay product energy density $\rho_\text{dr}$ as a function of the scale factor, all scaled by $a^3$. Evidently, the DWDM energy density redshifts as $\rho_H \propto a^{-4}$ while relativistic, turning non-relativstic around $a_\text{nr}\approx 3.15 T_0/m_H$ marked by the vertical dashed line in the figure, after which it converges to the rest mass energy and subsequently decays exponentially after a small period of $\rho_H \propto a^{-3}$ redshifting. The decay product energy density increases steadily during decay and simply redshifts as $a^{-4}$ after the decay is complete. Insofar as the decay process happens instantaneously, then, the evolution of the total energy density $\rho_\text{tot}$ of the decaying sector consists of three consecutive epochs:
\begin{itemize}
	\item $\rho_\text{tot} \propto a^{-4}$ while the decaying particle is relativistic.
	\item $\rho_\text{tot} \propto a^{-3}$ between the non-relativistic transition and the onset of decay. During this time, the energy of the decaying species is dominated by its rest mass energy $m_H n_H$.
	\item $\rho_\text{tot} \propto a^{-4}$ after the completion of the decay.
\end{itemize}
This scaling is characteristic of the warm dark matter decay. The middle section of $a^{-3}$ redshifting distinguishes it from a model of pure dark radiation, and the initial section of $a^{-4}$ redshifting distinguishes it from cold dark matter decay. The balance between the two first sections can be tuned by the mass and decay constant. Consequently, the DWDM model constitues a smooth interpolation between DCDM and added $N_\text{eff}$ models; a phenomenon we will see again later.

\subsection{Perturbations}
\begin{figure}[tb]
	\centering 
	\includegraphics[width=\textwidth]{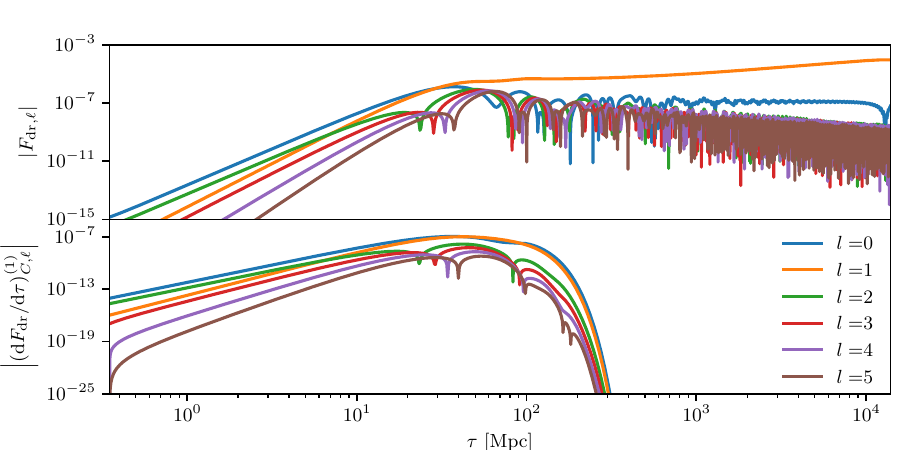}
	\caption{\label{fig6} Decay product perturbations in a run with $m_H = 10.0$ eV and $\Gamma=10^8$ km s$^{-1}$ Mpc$^{-1}$ at the scale $k=0.2$ Mpc$^{-1}$. \emph{Top panel:} Conformal time evolution of the first six Legendre components of the momentum averaged perturbations to the decay product distribution function. It is seen that the moments decrease with increasing $\ell$, with the exception of the $\ell=1$ multipole, which dominates at late times. \emph{Bottom panel:} Conformal time evolution of the Legendre components of the collision terms (\ref{col_int}). Apparently, it is fair to approximate their vanishing (at least for this specific set of parameter values) above some adequate $\ell$ value, but we still compute them in the following due to the computational efficiency of the recurrence relation computation based on (\ref{F_recurrence}).}
\end{figure}

In this subsection, we detail the numerical implementation of the perturbations and show sample solutions, starting with the decay products. The works~\cite{chacko2020, blinov2020, abellan2021i} set the dark radiation collision terms to zero for $\ell > 3$ since they are expensive to calculate directly and induce only a small error in the predicted CMB spectrum (although we find that it incurs sizable errors in species-specific quantities such as $\delta_\text{dr}$ and $\theta_\text{dr}$). However, with the recurrence relation (\ref{F_recurrence}), the computation time of the collision integrals is reduced immensely, and we can evaluate the full collision terms up to the maximum $\ell$ of the dark radiation hierarchy with only a very marginal increase in runtime. The first panel in figure~\ref{fig6} illustrates the momentum averaged decay product perturbations $F_\text{dr}$ for $k=0.2$ Mpc$^{-1}$ and $\ell$ values up to $\ell=5$ in a run with $m_H = 10.0$ eV and $\Gamma=10^8$ km s$^{-1}$ Mpc$^{-1}$. It is seen that the $\ell=1$ multipole overtakes the $\ell=0$ multipole and dominates around the non-relativistic transition and onwards, where the higher $\ell$ moments oscillate with an approximate period $2\pi /k$ and decay slowly. In the second panel, we show the collision terms (\ref{col_int}), computed with the fast recurrence relation. At all times, the magnitudes of the collision terms decrease drastically with $\ell$, and they all decay exponentially at the onset of the decay. Evidently, it is a decent approximation to truncate the collision term computation at some adequate $\ell$ value for this set of model parameters. This approximation is expected to become worse at small masses of the decaying particle, but we find that even for masses down to $0.01$\textendash$0.1$ eV, there is an appreciable gap between each subsequent $\ell$ moment of the collision term. Nonetheless, since the computation time is almost negligible, we still compute them for all $\ell$ values.

As for the decaying species, since its perturbed Boltzmann hierarchy (\ref{mother_hierarchy}) is identical to the hierarchy of a stable species in synchronous gauge, we evolve it like an ordinary non-cold species, but with a dynamical value of the logarithmic derivative $\d \ln \overline{f}_H / \d \ln q$ obtained from the background solution. By inspecting the hierarchy in synchronous gauge, it is seen that the latter always occurs as a front factor to the source terms from the metric. In particular, we find that $\d \ln \overline{f}_H / \d \ln q$ departs from its static Fermi-Dirac value, $-q\text{e}^q(1 + \text{e}^q)^{-1}$, by increasing gradually from large momenta downward. Since it is negative initially, the effect is a gradual reduction of its magnitude at large momenta, which manifests in a reduced sourcing from the metric terms. The physical interpretation of this phenomenon is that the population that at any point has not decayed yet will be increasingly dominated by large-$q$ particles which cluster less than their slower counterparts. In effect, the perturbations $\Psi_{H}(q)$ are reduced at large $q$ relative to the stable case.

\begin{figure}[tb]
	\centering 
	\includegraphics[width=\textwidth]{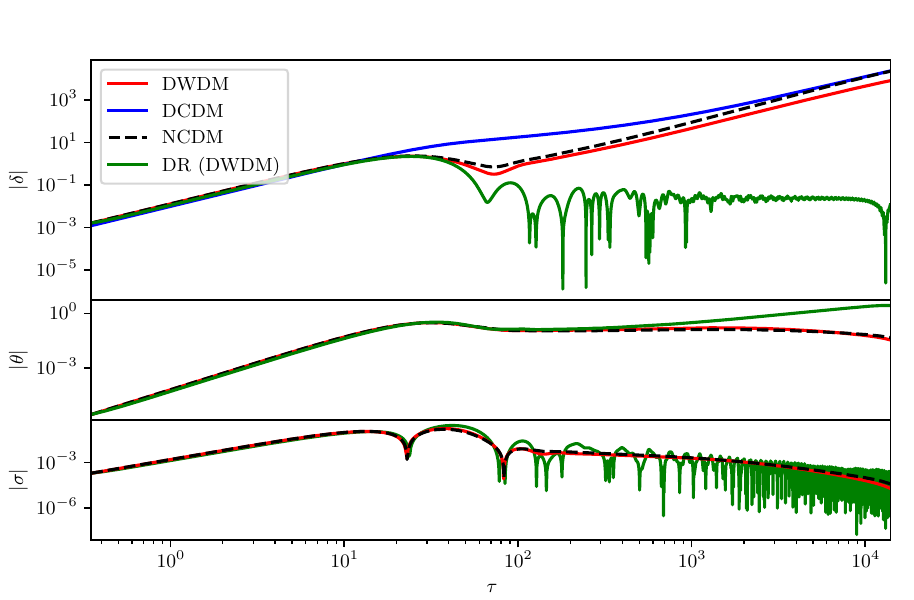}
	\caption{\label{fig5} Evolution of the integrated moments of the perturbed distribution functions of a variety of species for $k=0.2$ Mpc$^{-1}$ with conformal time. The DWDM and DCDM models share the decay constant $\Gamma = 10^8$ km s$^{-1}$ Mpc$^{-1}$ and have energy densities scaled such that their decay products contribute $\Delta N_\text{eff} = 0.5$ today. The stable non-cold species (NCDM) is given the same parameters as the DWDM species, e.g. a mass $m=10$ eV. It is seen that the DWDM species clusters less than both its stable counterpart and the cold decaying species.}
\end{figure}

The consequences this has for the integrated overdensity $\delta$, the velocity divergence $\theta$ and anisotropic stress $\sigma$ is shown on figure~\ref{fig5}, where the perturbations of an $m=10.0$ eV, $\Gamma = 10^8$ km s$^{-1}$ Mpc$^{-1}$ DWDM species, its decay product, a $\Gamma=10^8$ km s$^{-1}$ Mpc$^{-1}$ DCDM species and a $m=10.0$ eV stable non-cold species (NCDM) are compared. From examining the $\delta$ perturbations of the DWDM, DCDM and NCDM species in the figure and elsewhere, we report three features of the clustering of decaying species:
\begin{itemize}
	\item \emph{Decaying cold dark matter clusters exactly as much as stable cold dark matter}. This has been known for a long time~\cite{audren2014}, and we simply repeat it to set a context for the points below.
	\item \emph{Decaying warm dark matter clusters less than decaying cold dark matter}. This comes as no surprise, since it is well known that warm dark matter clusters less than cold dark matter, and the decay cannot affect this.
	\item \emph{Decaying warm dark matter clusters less than stable warm dark matter}. This final point, which is evident from figure~\ref{fig5}, is non-trivial, and is a consequence of the  mentioned fact that the large momentum part of the decaying population survives longest, so the mean momentum of the population increases as it undergoes decay. 
\end{itemize}
In the above, the first point is a special case of the third point: In the DCDM limit, the entire population decays at the same rate, so the mean momentum is constant, and the relative clustering mimics that of a stable species. Ultimately, therefore, the decaying warm species provides a structure formation pattern that generalizes both stable warm dark matter and decaying cold dark matter.

Reference~\cite{blinov2020} found that the relative overdensity $\delta$ of the decaying warm species converged to that of a decaying cold species. We have found that this behaviour results from not computing the physical densities beyond the point at which they become smaller than machine precision due to decay (which is remedied in the current work with the rescaling scheme described in section~\ref{sec:3.1}). Due to the rather long runtimes of the model, it is tempting to employ the standard fluid approximation for non-cold species in \textsc{class}~\cite{CLASS4}, however, one must be careful not to use the exact fluid approximation scheme for a stable species, since the continuity equation of the decaying sector deviates from the former. In appendix~\ref{apC}, we sketch how to correctly implement the fluid approximation for the decaying warm species. In particular, we find that it produces significantly erroneous values for the species specific perturbations (although only a negligible impact on the predicted CMB spectrum since the density parameter of the decaying sector is relatively small) whilst giving only a very marginal decrease in runtime. Accordingly, we refrain from using the fluid approximation in the rest of this work.

\subsection{Observable effects}
We shall now discuss the effects of the model on the cosmic microwave background spectrum and the matter power spectrum. We fix the following cosmological parameters: $\omega_b = 0.022382$, $\omega_\text{cdm} = 0.12010$, $A_s = 2.100549\cdot 10^{-9}$, $n_s = 0.966049$, $\tau_\text{reio} = 0.054308$ and chose to fix the acoustic scale $100 \theta_s = 1.042143$ instead of $H_0$, as in section~\ref{sec:1.1}, since the former is well constrained by CMB data. Furthermore, we adjust the energy densities of the models such that their contribution to the radiation energy density today is $\Delta N_\text{eff} = 0.5$. 

\begin{figure}[tb]
	\centering 
	\includegraphics[width=\textwidth]{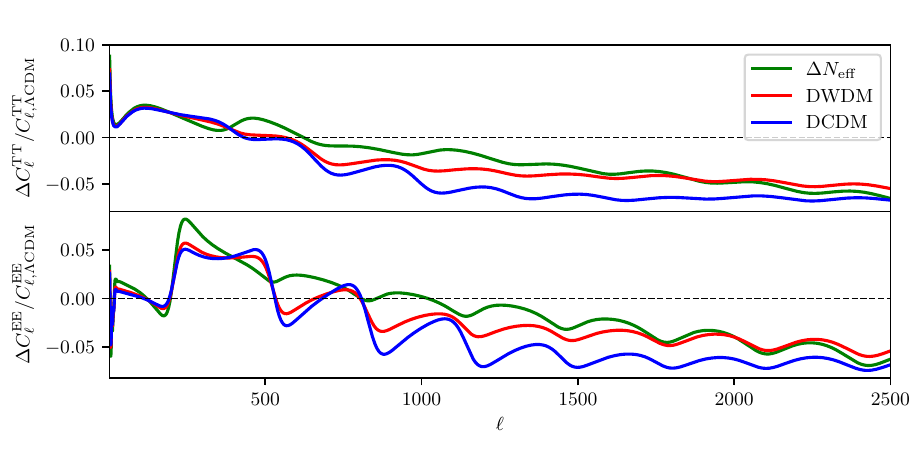}
	\caption{\label{fig7} CMB TT and EE spectrum relative to the $\Lambda$CDM spectrum for the $m=10$ eV DWDM model, a $\Delta N_\text{eff}$ model and a DCDM model. The decay constants of the decaying models is $\Gamma = 10^8$ km s$^{-1}$ Mpc$^{-1}$, and the models have been matched so as to contribute radiation energy density equivalent to $\Delta N_\text{eff} = 0.5$ today. This choice of parameters ensures a decay before recombination, and hence a substantial impact on the CMB spectrum.}
\end{figure}

Figure~\ref{fig7} illustrates the effect of the DWDM model on the CMB spectrum. The figure also shows the impact of a $\Delta N_\text{eff}$ model and a DCDM model with decay constant $\Gamma = 10^8$ km s$^{-1}$ Mpc$^{-1}$, which ensures that both decaying models have decayed before recombination. The effect on the CMB is small for decays that occur after recombination, so we do not show it. Intuitively, the DWDM effect on the CMB can be understood as a combination of the effects it inherits from the two other models, being the limiting cases for small and large masses. At large scales, there is an increased Integrated Sachs Wolfe effect~\cite{audren2014, poulin2016}; at small scales, there is a well-documented reduction in anisotropy characteristic of dark radiation~\cite{hou2011} present through its decay products. Altogether, the particular impact of DWDM is seen to interpolate between that of DCDM and $\Delta N_\text{eff}$, as expected.

\begin{figure}[tb]
	\centering 
	\includegraphics[width=\textwidth]{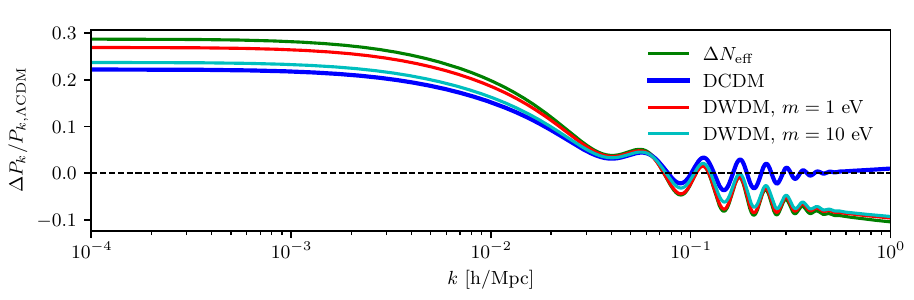}
	\caption{\label{fig8} Matter power spectrum for the DWDM model compared with $\Delta N_\text{eff}$ and DCDM models. The model parameters are the same as those in figure~\ref{fig7}.}
\end{figure}

Figure~\ref{fig8} shows the relative difference in matter power spectra $P(k)$ between the $\Lambda$CDM model and $m=1.0, 10.0$ eV DWDM models, a $\Delta N_\text{eff}$ dark radiation model and a DCDM model. As usual, the decaying models have the same decay constant, $\Gamma = 10^8$ km s$^{-1}$ Mpc$^{-1}$, and are matched to contribute the same radiation density $\Delta N_\text{eff} = 0.5$ at late times. Clearly, especially at large scales, the DWDM models interpolate between the DCDM and $\Delta N_\text{eff}$ models. The overall trend is that all models increase large scale power. The DCDM model increases small-scale power due to the strong small-scale clustering of cold dark matter\footnote{This effect naturally opposes the tendency of the relativistic decay products to reduce small-scale structure. With the model parameters chosen here, the cold dark matter clustering wins slightly, but e.g. in~\cite{audren2014, poulin2016}, small-scale structure is also reduced within the DCDM model.}, whereas the DWDM and $\Delta N_\text{eff}$ models suppress small-scale clustering, which can be understood by an argument similar to their suppression of small-scale anisotropies~\cite{hou2011}. Note that the power in the $m=1.0$ eV DWDM model approaches the $\Delta N_\text{eff}$ values at large scales, while the $m=10.0$ eV DWDM model approaches that of DCDM at large scales, but at small scales, they both lean towards the $\Delta N_\text{eff}$ value. This is most likely due to a reduction in power when the decaying species is relativistic.

%%%%%%%%%%%%%%%%%%%%%%%%%%%%%%%%%%%%%%%%%%%%%%%%%%%%%%%%%%%%%%%%%%%%%%%%
\section{Parameter constraints} \label{sec:4}
In this section, we conduct MCMC analyses to obtain posterior distributions for the parameters of the DWDM model. The implementation detailed in the last section includes a set of precision settings, such as the number of momentum bins of the decaying species and the multipoles where the Boltzmann hierarchies are truncated. Increasing these yield more accurate results at the expensive of increasing the execution time. Since a single computation can take anywhere between a few seconds and many minutes depending on these, we have chosen $13$ momentum bins and truncate the hierarchies at $\ell_{\text{max}} = 13$, for a runtime of $\approx 15$ seconds per model (on the $8$ cores of the Apple M1 Pro) and a maximum error in the predicted $C_\ell$'s around $1 \%$.

We compute posterior distributions with likelihoods based on two datasets. Our baseline $\mathcal{D}_{\text{base}}$ consists of the following data sets:
\begin{itemize}
	\item Planck 2018 (including high-$\ell$ TTTEEE, low-$\ell$ TT, EE and lensing)~\cite{planck2018},
	\item BAO (including BOSS DR12~\cite{boss2016} and low redshift data from the 6dF survey~\cite{beutler2011} and the BOSS main galaxy sample~\cite{ross2014}).
\end{itemize}
For the second set of likelihoods, we add supernova data from the Pantheon compilation~\cite{scolnic2017} in order to confront the model with local Universe observations. Together, these two datasets correspond to the first two tests in the comparison of proposed tension solutions in reference~\cite{schoeneberg2021}. 
When employing Pantheon data and comparing with the local SH0ES $H_0$ measurement, one should be careful not to alter the late Universe dynamics affecting the calibration of the SNIa data, as discussed in reference~\cite{benevento2020}; if this is done, the most correct approach is to use the calibration of the intrinsic SNIa magnitude $M_b$ as the target observable~\cite{benevento2020, camarena2021}. However, for all interesting areas in parameter space, the DWDM model does not introduce radical changes to the luminosity distance at the small redshifts relevant to the calibration, so we keep $H_0$ as the tension target also when including Pantheon data. 

As for the cosmological parameters scanned over, those pertaining to the decaying species are detailed in the next subsection; otherwise we take the usual set of $\Lambda$CDM parameters
\begin{align} \nonumber
	\left\{ \omega_\text{b}, \omega_\text{cdm}, H_0, \ln 10^{10} A_s, n_s, \tau_\text{reio} \right\}.
\end{align}
In particular, for each of the two dataset combinations described above, we have run six Markov chains using the \textsc{MontePython} code~\cite{Audren:2012wb, montepython} and checked for convergence both through a Gelman-Rubin criterion of $R-1 \lesssim 0.05$ and by ensuring that the posteriors only vary negligibly with additional running time. The complete two and one-dimensional marginalized posterior distributions can be seen as triangle plots in appendix~\ref{apD}, and the resulting parameter constraints are summarized in table~\ref{tab:full}, also in appendix~\ref{apD}.

\subsection{Decay parameters}
In this section, we present and discuss posterior distributions focusing on the parameters specific to the DWDM model, namely the initial density, parametrized as $N_{\text{eff,ini,dwdm}}$, the lifetime $\tau$ and the decaying particle mass, $m$\footnote{We note that the decaying species is added on top of a fixed amount of dark radiation and a single massive neutrino species corresponding to $N_{\text{eff}} = 3.046$, in accordance with the recommendation from Planck measurements~\cite{planck2018}.}. The priors we have chosen are
\begin{figure}[tb]
	\centering 
	\includegraphics[width=\textwidth]{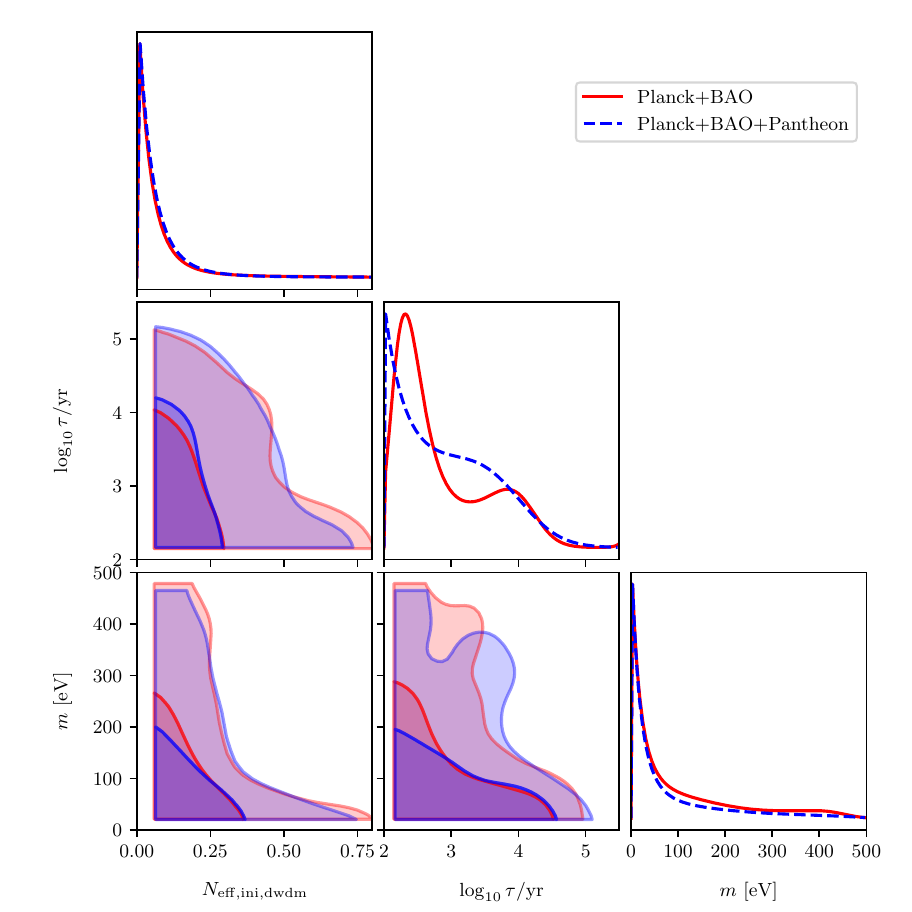}
	\caption{\label{fig_triangle} Marginalized posterior distributions for the three cosmological parameters introduced with the warm decaying species using two datasets as described in the text. Note the bump on the lifetime posterior, which is characteristic of decaying dark matter models; it is also seen in the DCDM model~\cite{nygaard2020} and in the DWDM model of~\cite{blinov2020}. Note also that the peak in $\tau$ for Planck+BAO is an artefact of the binning used in the analysis of the Markov chain; its posterior actually peaks at the lower prior bound like it does with the added Pantheon data.}
\end{figure}
\begin{align} \nonumber
	N_{\text{eff,ini,dwdm}} \in [0, 3], \quad \log_{10} \tau /\text{yr} \in [2, 6], \quad m / \text{eV} \in [0.001, 500].
\end{align}
While the $N_{\text{eff,ini,dwdm}}$ and $m$ priors are fairly generous, the lifetime prior lies at somewhat small values of $\tau$, roughly corresponding to the short-lived regime of the DCDM model in references~\cite{nygaard2020, poulin2016}. Although we expect the very long-lived region of parameter space to also be viable, small lifetimes constitute the relevant regime for addressing the $H_0$ tension~\cite{nygaard2020}, and ultimately, this regime is not necessarily short-lived for a DWDM species due to time dilation of the lifetime. 
\begin{table}[tb]
	\centering
	\begin{tabular}{? Sc | Sc  Sc  Sc ?}
		%\hline
		\specialrule{.12em}{0em}{0em} 
		\bf Data & $N_{\text{eff,ini,dwdm}}$ & $\log_{10} \tau /\text{yr}$ & $m\times 10^2$ [eV] \\
		\hline
		Planck$+$BAO & $<0.105$ & $<3.06$ & $<1.23$  \\
		Planck$+$BAO$+$Pantheon & $<0.109$ & $<3.29$ & $<0.82$ \\
		& & & \\
		\specialrule{.12em}{0em}{0em} 
	\end{tabular}
	\caption{\label{tab:constraints} Constraints on DWDM parameters derived from MCMC analyses described in the text. The uncertainties indicate $1\sigma$ intervals, corresponding to a $68 \%$ confidence level. As explained in the text, the $\log_{10} \tau$ constraints are prior dependent and therefore not directly meaningful.}
\end{table}
Figure~\ref{fig_triangle} illustrates posteriors for and correlations between the aforementioned parameters specific to the DWDM model. First and foremost, we observe that all one-dimensional posteriors obtain their maximum values at an endpoint of the prior range (with the exception of the Planck+BAO lifetime due to poor binning). Thus, we find no detection of a decaying warm species, although the data does admit a modest component of DWDM. All constraints on these parameters are therefore upper bounds, and can be found in table~\ref{tab:constraints}. From conducting additional small MCMC runs with various lifetime priors, we find that the posterior always obtains its maximum at the lower lifetime bound. Decreasing the lower prior bound therefore shifts the obtained bounds, so we can obtain no meaningful upper bound on $\log_{10} \tau$. The other parameters evade this issue since they have a physically motivated lower bound of zero; however, due to the long tail in the $m$ posterior, we also expect our constraint on the mass to vary slightly with the upper prior bound (for example, reference~\cite{abellan2021neutrino} use a physically motivated upper bound on the mass prior and thereby find much tighter constraints on $m$). 

Another issue inherent in the DWDM posterior comes from a volume effect: When the abundance $N_\text{eff,ini,dwdm}$ approaches $0$, the lifetime and mass parameters must become unconstrained, giving a significant boost to the posterior volume around $N_\text{eff,ini,dwdm}\approx 0$. As a consequence, when marginalized over the lifetime and mass, the posterior will unfairly favour the $N_\text{eff,ini,dwdm}\approx 0$ region\footnote{This phenomenon is common in $\Lambda$CDM extensions since they must, by definition, include some abundance parameter such that any additional model parameters become unconstrained at the vanishing of the former. Early dark energy models are a typical example; for an investigation using a profile likelihood, see reference~\cite{Herold:2021ksg}.}. Profile likelihood methods have proven very succesful at evading such volume effects (e.g.~\cite{Gomez-Valent:2022hkb, Herold:2021ksg, Hamann_2012}), but we leave a further investigation of the consequence for the DWDM model open to future work.

\begin{figure}[tb]
	\centering 
	\includegraphics[width=\textwidth]{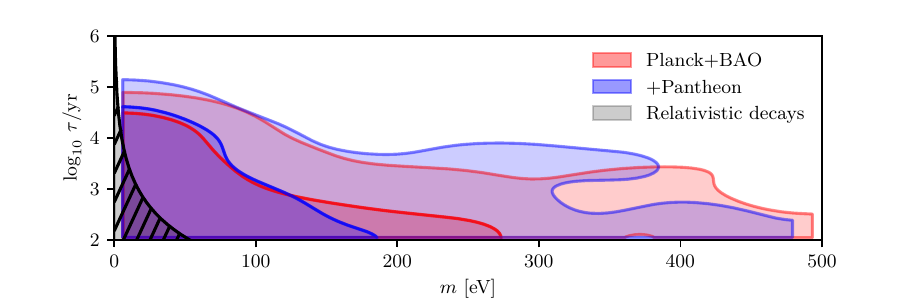}
	\caption{\label{fig_relativistic} Two-dimensional marginalized posterior over the decaying species mass $m$ and lifetime $\tau$ for the baseline dataset Planck+BAO (red) and the baseline including Pantheon data (blue). In the scratched area, the decays are relativistic, and inverse decay processes must be modelled for an accurate description of the physics. We have defined this region as that below the $\alpha=1$ line, with $\alpha$ denoting the relativity parameter (\ref{relativity_parameter}).}
\end{figure}

On figure~\ref{fig_relativistic}, the two-dimensional mass\textendash lifetime posterior distributions are shown again, along with the region $\alpha < 1$, where $\alpha$ is the relativity parameter introduced in equation (\ref{relativity_parameter}), and we recall that relativistic decays correspond to the region $\alpha < 1$. It is seen that an appreciable area of the posterior volume is contained in the regime of relativistic decays; especially if one extrapolates to smaller lifetimes. Moreover, the maximum of the posterior lies deep in the area of relativistic decays. Since our model is not physically meaningful in this regime, one could exclude it by employing a prior corresponding to the region of non-relativistic decays, as was done in reference~\cite{abellan2021neutrino}. However, in order to properly investigate whether the apparent favorization of the relativistic regime is an artefact of the current model's inability to describe the physics or if it is actually favoured by data, a complete implementation of the model including inverse decays, and possibly quantum statistical effects, should be developed. We leave this opportunity for future work.

\subsection{$H_0$ and $\sigma_8$ tensions}
In this section, we study the impact of the DWDM model on the value of the Hubble constant $H_0$ and $\sigma_8$ and assess to what extent it is able to alleviate the associated cosmological tensions.

\begin{figure}[tb]
	\centering 
	\includegraphics[width=\textwidth]{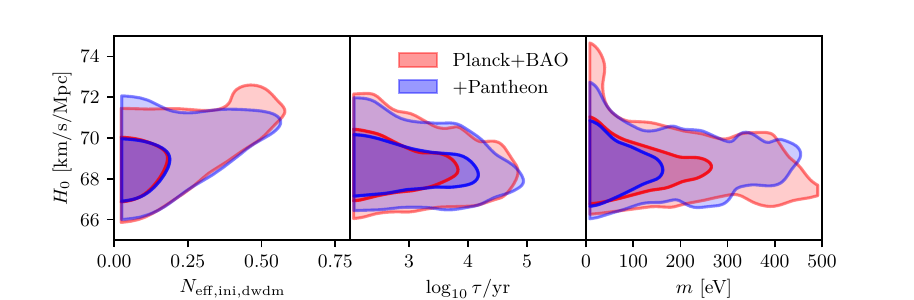}
	\caption{\label{fig_hubble_contours} Two-dimensional posterior distributions of $H_0$ and the three DWDM model parameters described in the text for two datasets.}
\end{figure}

Firstly, we highlight the impact of each of the three model parameters $N_{\text{eff,ini,dwdm}}$, $\log_{10} \tau$ and $m$ on the marginalized $H_0$ posterior in the two-dimensional posteriors shown in figure~\ref{fig_hubble_contours}. The initial energy density, parametrized as $N_{\text{eff,ini,dwdm}}$, correlates positively with $H_0$. This can be understood from the arguments evoked in section~\ref{sec:1.1}: Additional early radiation increases $H(z)$ before recombination, which requires $H(z)$ to increase after recombination in order to fix the CMB peak position enforced model independently by observations. Within the uncertainties, the lifetime is seem to be largely uncorrelated with $H_0$. The decaying species mass $m$ also seems to be rather uncorrelated with $H_0$, although the $H_0$ posterior widens at smaller masses\footnote{Reference~\cite{blinov2020} found that large masses yielded slightly larger best-fit $H_0$ values; however, this correlation was not significant compared to the uncertainties in the analysis. References~\cite{Escudero:2021rfi, abellan2021neutrino} also only find very weak correlations between $m$ and $H_0$.}. Actually, the lack of correlation between $m$ and $H_0$ gives an interesting corollary, namely that the limiting models of DCDM and $\Delta N_\text{eff}$ should approximately achieve the same best-fit value of $H_0$, a result that is more or less corroborated by the current literature~\cite{nygaard2020, schoeneberg2021, divalentino2021}.
\begin{figure}[tb]
	\centering 
	\includegraphics[width=\textwidth]{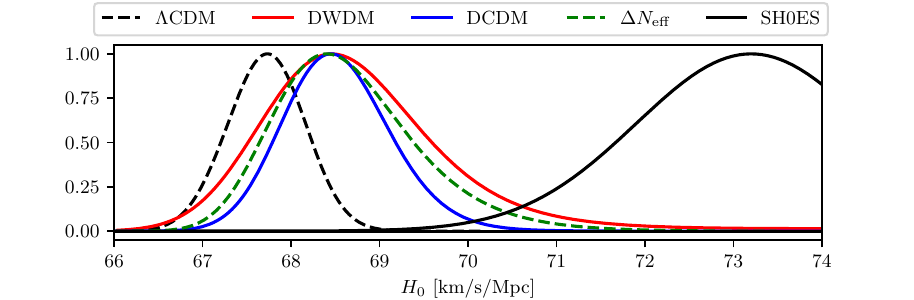}
	\caption{\label{fig_hubble} One-dimensional marginalized posterior distribution for $H_0$ for different cosmological models: A $\Lambda$CDM model, a DWDM model with, a short-lived DCDM model and the value from the local measurement by the SH0ES collaboration~\cite{riess2021}. The DWDM, \text{dcdm} and $\Delta N_\text{eff}$ models are seen to display similar predictions for $H_0$.}
\end{figure}

To illuminate this conclusion further, figure~\ref{fig_hubble} shows the one-dimensional marginalized posterior distributions for $H_0$ for a $\Lambda$CDM model, a DWDM model where the mass is marginalized away, DCDM and $\Delta N_\text{eff}$ models as well as the local $H_0$ measurement from the SH0ES collaboration. It is immediately seen that all of the DWDM, DCDM and $\Delta N_\text{eff}$ models admit larger values of $H_0$ than $\Lambda$CDM. Furthermore, the posteriors of these models peak at the same $H_0$ value, with DWDM and $\Delta N_\text{eff}$ having larger widths than DCDM, all as expected from the non-correlation between $H_0$ and the mass parameter of DWDM just discussed.

As a concrete quantitative statistic illuminating the alleviation of the tensions and the quality of the fit to data, we employ the Gaussian tension between the posteriors on the parameters $x_\mathcal{D}\in \{H_0, \sigma_8\}$ and reference values $x_\text{ref}$, defined as~\cite{raveri2018},
\begin{align} \nonumber
	\text{GT}(x_\mathcal{D}) = \frac{\overline{x}_\mathcal{D} - \overline{x}_{\text{ref}}}{\sqrt{\sigma_\mathcal{D}^2 + \sigma_{\text{ref}}^2}},
\end{align}
where $\overline{x}$ and $\sigma$ denote mean values and standard deviations, respectively. To this end, we employ the concrete value~\cite{riess2021}
\begin{align} \nonumber
	H_0 = 73.2 \pm 1.3 \text{ km s}^{-1}\text{ Mpc}^{-1},
\end{align}
\begin{table}[tb]
	\centering
	\begin{tabular}{? Sc | Sc  Sc | Sc | Sc ?}
		%\hline
		\specialrule{.12em}{0em}{0em} 
		\bf Data & $H_0$ [km s$^{-1}$ Mpc$^{-1}$] & GT($H_0$) & $\Delta \chi^2 (\mathcal{D}_{\text{base}})$ & $\Delta \chi^2(\mathcal{D}_{\text{base}}+\text{SH0ES})$ \\
		\hline
		Planck$+$BAO & $68.73_{-1.3}^{+0.81}$ & $ 2.7\sigma $ & $0.22$ & $-6.18$ \\
		Planck$+$BAO$+$Pantheon & $68.65_{-1.2}^{+0.83}$ & $2.8\sigma$ & $-1.34$ & $-6.08$ \\
		& & & & \\
		\specialrule{.12em}{0em}{0em} 
	\end{tabular}
	\caption{\label{tab:tensions} Results for the DWDM model from the MCMC runs described in the text. The last columns represent the difference in $\chi^2$ values for the DWDM and $\Lambda$CDM models at their best-fit points, $\Delta \chi^2 = \chi^2_\text{min,DWDM} - \chi^2_{\text{min,}\Lambda\text{CDM}}$.}
\end{table}
which is at a $4.1\sigma$ Gaussian tension with the value inferred from the Planck collaboration, $H_0 = 67.27 \pm 0.60$ km s$^{-1}$ Mpc$^{-1}$~\cite{planck2018}. The Gaussian tension fails as a measure of the tension when the model posterior departs from Gaussianity. This can be generalized using the difference of maximum a posteriori metric $Q_\text{DMAP}$ from reference~\cite{raveri2018}, but since we find mainly Gaussian one-dimensional $H_0$ posteriors, we refrain from computing this. The results are presented in table~\ref{tab:tensions}, where it is seen that the $H_0$ tension is alleviated by $1.3\sigma$ and $1.4\sigma$ with and without Pantheon data, respectively. Since this is a modest alleviation, we conclude that the non-relativistically decaying DWDM model cannot resolve the Hubble tension. In order to also quantify the quality of the fit to the entire dataset, we compute the difference in $\chi^2$ values at the best-fit points between the DWDM model and the $\Lambda$CDM model, also given in the table\footnote{Sometimes the difference in $\chi^2$ values is incremented by double the amount of extra parameters in the theory such that it becomes the difference in Akaike Information criteria (AIC)~\cite{akaike1974}. We do not use this here since, for example, the penalty of having the mass parameter could be avoided by fixing it to its best fit, equivalent to a model of pure radiation.}. For Planck and BAO data only, we find that the DWDM model is as good a fit as $\Lambda$CDM, which matches the general result that the posterior maximum lies in the $\Lambda$CDM limit. Including Pantheon data slightly increases the goodness of fit of DWDM, and including a Gaussian likelihood on the SH0ES value of $H_0$ significantly increases the preference of DWDM. Since MCMC methods are very inefficient at finding best-fit points~\cite{Hamann_2012}, we have used a simulated annealing approach, based on reference~\cite{Hannestad:2000wx}, as the optimization algorithm. We generally find an improvement of around $2$\textendash$5$ $\chi^2$ degrees of freedom relative to the MCMC best-fit with this approach and assess the uncertainty to be around $0.5$\textendash$1$ $\chi^2$ degrees of freedom.
\begin{table}[tb]
	\centering
	\begin{tabular}{? Sc | Sc  Sc | Sc Sc | Sc ?}
		%\hline
		\specialrule{.12em}{0em}{0em} 
		\bf Model & $H_0$ [km s$^{-1}$ Mpc$^{-1}$] & GT($H_0$) & $S_8$ & GT($S_8$) & $\Delta \chi^2$ \\
		\hline
		DCDM &  $68.64_{-0.81}^{+0.45}$ & $3.2\sigma$ & $0.828_{-0.018}^{+0.016}$ & $2.2 \sigma$ & $0.68$ \\
		DWDM & $68.73_{-1.3}^{+0.81}$ & $2.7\sigma$ & $0.825_{-0.014}^{+0.014}$ & $2.2\sigma$ & $0.22$ \\
		$\Delta N_\text{eff}$ & $68.66_{-1.0}^{+0.63}$ & $3.0\sigma$ & $0.825_{-0.011}^{+0.011}$ & $2.4\sigma$ & $0.34$  \\
		& & & & &  \\
		\specialrule{.12em}{0em}{0em} 
	\end{tabular}
	\caption{\label{tab:comp} Planck 2018 + BAO comparison of the DWDM model and its two limits, the (short-lived) DCDM model and a model with pure additional dark radiation parametrized as $\Delta N_\text{eff}$, in terms of their ability to alleviate the $H_0$ and $S_8$ tensions, and their overall fit to the data, $\Delta \chi^2 = \chi^2_\text{min} - \chi^2_{\text{min,}\Lambda\text{CDM}}$.}
\end{table}
Table~\ref{tab:comp} provides the same statistics but for a fixed Planck+BAO dataset and for the DWDM model as well as its two limiting models, the decaying cold dark matter and pure, invisible radiation $\Delta N_\text{eff}$. As mentioned, the $H_0$\textendash mass contour in figure~\ref{fig_hubble_contours} indicates that the mass and Hubble constant are largely uncorrelated. Since the mass is the parameter that interpolates the DWDM model between its limits, one therefore expects the DCDM and $\Delta N_\text{eff}$ models to predict similar values for $H_0$, and indeed, as can be seen from the results in the table, this is what we find. Furthermore, since the best-fit points found for the three models lie in the $\Lambda$CDM limit, the minimum $\chi^2$ values are identical up to uncertainties.

\begin{figure}[tb]
	\centering 
	\includegraphics[width=\textwidth]{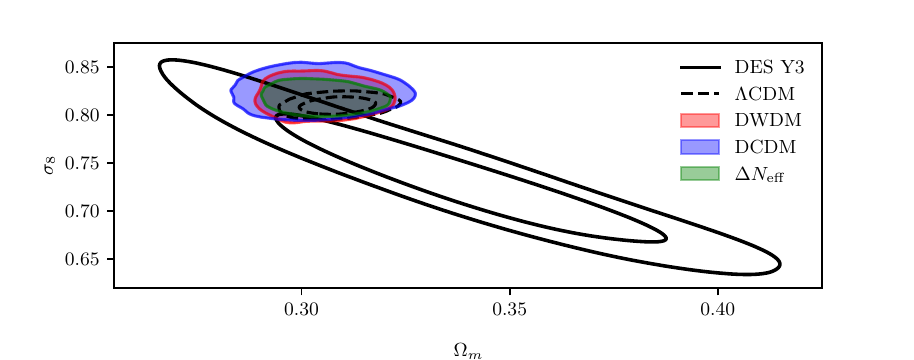}
	\caption{\label{fig_sigma} Two-dimensional marginalized posteriors of the current matter density parameter $\Omega_m$ and $\sigma_8$ for the $\Lambda$CDM model, the locally measured posterior from the fiducial $3\times 2$pt year 3 data release of the DES collaboration~\cite{DES} as well as the DWDM model, the short-lived DCDM model, and a $\Delta N_\text{eff}$ model. To reduce clutter, only the $2\sigma$ contours, corresponding to a 95 \% confidence level, are shown for the latter three models.}
\end{figure}

To assess the impact of the DWDM model on the $S_8$ tension, figure~\ref{fig_sigma} illustrates the $(\Omega_m, \sigma_8)$ posteriors of the DWDM model and its two limiting models as well as the $\Lambda$CDM model and that obtained from the recent year 3 data release of the Dark Energy Survey (DES) collaboration~\cite{DES}. It is seen that the $\Lambda$CDM, DWDM, DCDM and $\Delta N_\text{eff}$ models share a lower bound on $\sigma_8$, corresponding to the $\Lambda$CDM limit, and the three latter models all admit larger upper bounds on $\sigma_8$. Contrary to several of the best proposed solutions to the $H_0$ tension, we do not expect the DWDM model and its limits to worsen the tension in $\sigma_8$ appreciably.

Since the $\sigma_8$ posterior is highly non-Gaussian, as seen on figure~\ref{fig_sigma}, the Gaussian tension is not an appropriate measure of the discrepancy with the value from the DES collaboration. We therefore reparameterize the $\sigma_8$ parameter as $S_8 \equiv \sigma_8 (\Omega_m /0.3)^{0.5}$, yielding a fairly Gaussian posterior~\cite{DES}. The level of the tension is then estimated using the mean value and $68$ \% confidence limits of the recommended fiducial $3\times 2$ pt value of $S_8$ obtained by the DES collaboration,
\begin{align}
	S_8 \equiv \sigma_8 (\Omega_m / 0.3)^{0.5} = 0.776 \pm 0.017,
\end{align}
which is at a $2.5\sigma$ Gaussian tension with the value from CMB measurements by Planck, $S_8 = 0.834\pm 0.016$~\cite{planck2018}. The corresponding values for the DCDM, DWDM and $\Delta N_\text{eff}$ models, along with the resulting Gaussian tension measures, are shown in table~\ref{tab:comp}. These numbers corroborate the conclusion from figure~\ref{fig_sigma} that the models neither alleviate the $S_8$ tension nor worsen it. It was shown in reference~\cite{abellan2021i} that a decaying cold dark matter model with massive decay products could alleviate the $S_8$ tension, which can be understood as a consequence of the finite free-streaming length of the massive decay products. Since the DWDM model is a generalization of decaying cold dark matter, we also expect the DWDM model to be able to alleviate the $S_8$ tension if one allows for massive decay products. In this case, the DWDM model becomes one of only few models to actually help \emph{both} the $H_0$ and $S_8$ tensions. We leave the study of decaying warm dark matter with massive decay products for future work.

%%%%%%%%%%%%%%%%%%%%%%%%%%%%%%%%%%%%%%%%%%%%%%%%%%%%%%%%%%%%%%%%%%%%%%%%
\section{Conclusion} \label{sec:5}
In this work, we have performed a comprehensive study of non-relativistically decaying warm dark matter (DWDM) with dark radiation decay products. There exist several realistic particle physics realizations of the model~\cite{blinov2020}, and we have shown explicitly how it arises from a universal interaction between two neutrino-like species and a light scalar particle based on reference~\cite{barenboim2020}. A key feature is that its lifetime is time dilated due to the non-negligible momentum of the species, resulting in delayed decays compared to a decaying cold dark matter species. Interestingly, this characteristic causes the at any point surviving population to become increasingly dominated by particles of large comoving momenta, which diminishes their tendency to cluster relative to the corresponding stable species. 

The background evolution of the decaying sector can largely be grouped into an early, relativistic epoch, followed by an intermediate period where the species has become non-relativistic but has not yet decayed, and then finally an epoch after the decay where the sector again redshifts like radiation due to the energy deposited in the dark radiation decay products. This flexible juggling of equation of states in the decaying sector provides substantial freedom in its impact on the expansion history. An approximate analytical solution to the background equation, valid to about $10 \%$, is presented, which is useful for brief estimates of the time evolution of the species. Moreover, we derived a recurrence relation, in the multipole order, of the decay kernel $\mathcal{F}_\ell (x)$ that appears in the collision term of the momentum averaged decay product perturbations. With this, the computation of the decay product collision term becomes a strongly sub-dominant contribution to the total computation time in the Einstein-Boltzmann solver, although the impacts of the collision term on observables such as the $C_\ell$'s rapidly become negligible with increasing multipole.

Since the decaying species is relativistic in the early Universe, it contributes additional radiation energy density and increases the Hubble parameter at early times. As a consequence, the Hubble parameter today is increased in order to anchor the acoustic scale at recombination to observations. With this motivation, we have conducted MCMC analyses in order to investigate the ability of the warm decaying model to alleviate the Hubble and $S_8$ tensions, and find a rather mild alleviation of around $\sim 1$\textendash$2\sigma$ for $H_0$ and no $S_8$ alleviation, using Planck 2018 as well as BAO and Pantheon data. On the one hand, the DWDM species converges to a decaying cold dark matter (DCDM) species in the limit of large masses. On the other hand, it converges to pure dark radiation in the limit of small masses, since the decay then becomes kinematically unfeasible. Consequentially, the DWDM species interpolates between the DCDM and dark radiation models as a function of its mass, which is also evident from its effects on the CMB and matter power spectra. Interestingly, we find $H_0$ to be largely uncorrelated with the DWDM mass. Since the latter interpolates the model between DCDM and dark radiation, we obtain as a corollary that DCDM and $\Delta N_\text{eff}$ models should have similar impacts on the Hubble tension, which we show is corroborated by data. With the MCMC analyses, we find that a modest population of a DWDM species is compatible with data. Furthermore, data prefers small masses and lifetimes, corresponding to a fast decay and convergence to a model of dark radiation. However, in this area of parameter space, the DWDM particle decays while still relativistic, such that inverse decays and their quantum statistical corrections become important~\cite{barenboim2020}. In order to properly establish the complete behaviour of the DWDM model and its relation to observational data, then, these processes should be taken into account. Since this is a difficult and expensive numerical undertaking, we leave it open for future work. \\

\noindent \textbf{Reproducibility.} The code used to obtain the results in this paper is available at \url{https://github.com/AarhusCosmology/CLASSpp\_public} on the branch \textsc{2205.13628} and SHA 03be0ef1e0f8b5bacce975eb9e58661e4d9a7e5f. The version of \textsc{MontePython} 3.5 used, as well as parameter files and plotting scripts are available at \url{https://github.com/AarhusCosmology/montepython\_public} on the branch \textsc{2205.13628}. \\

%\section*{Acknowledgements}
\noindent \textbf{Acknowledgements.} The authors are very grateful to Nikita Blinov for useful discussions and interpretations of our results. The numerical computations presented in this work were conducted at the Centre for Scientific Computing, Aarhus \url{https://phys.au.dk/forskning/faciliteter/cscaa}. E.B.H. and T.T. are supported by a research grant (29337) from the VILLUM FONDEN.

%%%%%%%%%%%%
\bibliographystyle{utcaps}
%\nocite{*}
\bibliography{paper}

\providecommand{\href}[2]{#2}\begingroup\raggedright\begin{thebibliography}{10}

\bibitem{freedman2019}
W.~L. Freedman {\em et al.}, ``{The Carnegie-Chicago Hubble Program. VIII. An
  Independent Determination of the Hubble Constant Based on the Tip of the Red
  Giant Branch},'' \href{http://arxiv.org/abs/1907.05922}{{\ttfamily
  arXiv:1907.05922 [astro-ph.CO]}}.

\bibitem{divalentino2021}
E.~Di~Valentino, O.~Mena, S.~Pan, L.~Visinelli, W.~Yang, A.~Melchiorri, D.~F.
  Mota, A.~G. Riess, and J.~Silk, ``{In the Realm of the Hubble tension $-$ a
  Review of Solutions},'' \href{http://arxiv.org/abs/2103.01183}{{\ttfamily
  arXiv:2103.01183 [astro-ph.CO]}}.

\bibitem{planck2018}
{\bfseries Planck} Collaboration, N.~Aghanim {\em et al.}, ``{Planck 2018
  results. VI. Cosmological parameters},''
  \href{http://dx.doi.org/10.1051/0004-6361/201833910}{{\em Astron. Astrophys.}
  {\bfseries 641} (2020)  A6},
  \href{http://arxiv.org/abs/1807.06209}{{\ttfamily arXiv:1807.06209
  [astro-ph.CO]}}.

\bibitem{riess2021}
A.~G. Riess, S.~Casertano, W.~Yuan, J.~B. Bowers, L.~Macri, J.~C. Zinn, and
  D.~Scolnic, ``{Cosmic Distances Calibrated to 1\% Precision with Gaia EDR3
  Parallaxes and Hubble Space Telescope Photometry of 75 Milky Way Cepheids
  Confirm Tension with $\Lambda$CDM},''
  \href{http://dx.doi.org/10.3847/2041-8213/abdbaf}{{\em Astrophys. J. Lett.}
  {\bfseries 908} (2021) no.~1, L6},
  \href{http://arxiv.org/abs/2012.08534}{{\ttfamily arXiv:2012.08534
  [astro-ph.CO]}}.

\bibitem{heymans2020}
C.~Heymans {\em et al.}, ``{KiDS-1000 Cosmology: Multi-probe weak gravitational
  lensing and spectroscopic galaxy clustering constraints},''
  \href{http://dx.doi.org/10.1051/0004-6361/202039063}{{\em Astron. Astrophys.}
  {\bfseries 646} (2021)  A140},
  \href{http://arxiv.org/abs/2007.15632}{{\ttfamily arXiv:2007.15632
  [astro-ph.CO]}}.

\bibitem{schoeneberg2021}
N.~Sch\"oneberg, G.~Franco~Abell\'an, A.~P\'erez~S\'anchez, S.~J. Witte,
  V.~Poulin, and J.~Lesgourgues, ``{The $H_0$ Olympics: A fair ranking of
  proposed models},'' \href{http://arxiv.org/abs/2107.10291}{{\ttfamily
  arXiv:2107.10291 [astro-ph.CO]}}.

\bibitem{perivolaropoulos2021}
L.~Perivolaropoulos and F.~Skara, ``{Challenges for $\Lambda$CDM: An update},''
  \href{http://arxiv.org/abs/2105.05208}{{\ttfamily arXiv:2105.05208
  [astro-ph.CO]}}.

\bibitem{knox2019}
L.~Knox and M.~Millea, ``{Hubble constant hunter\textquoteright{}s guide},''
  \href{http://dx.doi.org/10.1103/PhysRevD.101.043533}{{\em Phys. Rev. D}
  {\bfseries 101} (2020) no.~4, 043533},
  \href{http://arxiv.org/abs/1908.03663}{{\ttfamily arXiv:1908.03663
  [astro-ph.CO]}}.

\bibitem{Nunes:2021ipq}
R.~C. Nunes and S.~Vagnozzi, ``{Arbitrating the S8 discrepancy with growth rate
  measurements from redshift-space distortions},''
  \href{http://dx.doi.org/10.1093/mnras/stab1613}{{\em Mon. Not. Roy. Astron.
  Soc.} {\bfseries 505} (2021) no.~4, 5427--5437},
  \href{http://arxiv.org/abs/2106.01208}{{\ttfamily arXiv:2106.01208
  [astro-ph.CO]}}.

\bibitem{poulin2016ii}
V.~Poulin, J.~Lesgourgues, and P.~D. Serpico, ``{Cosmological constraints on
  exotic injection of electromagnetic energy},''
  \href{http://dx.doi.org/10.1088/1475-7516/2017/03/043}{{\em JCAP} {\bfseries
  03} (2017)  043}, \href{http://arxiv.org/abs/1610.10051}{{\ttfamily
  arXiv:1610.10051 [astro-ph.CO]}}.

\bibitem{yuksel2007}
H.~Yuksel and M.~D. Kistler, ``{Circumscribing late dark matter decays model
  independently},'' \href{http://dx.doi.org/10.1103/PhysRevD.78.023502}{{\em
  Phys. Rev. D} {\bfseries 78} (2008)  023502},
  \href{http://arxiv.org/abs/0711.2906}{{\ttfamily arXiv:0711.2906
  [astro-ph]}}.

\bibitem{zhang2007}
L.~Zhang, X.~Chen, M.~Kamionkowski, Z.-g. Si, and Z.~Zheng, ``{Constraints on
  radiative dark-matter decay from the cosmic microwave background},''
  \href{http://dx.doi.org/10.1103/PhysRevD.76.061301}{{\em Phys. Rev. D}
  {\bfseries 76} (2007)  061301},
  \href{http://arxiv.org/abs/0704.2444}{{\ttfamily arXiv:0704.2444
  [astro-ph]}}.

\bibitem{scherrer1985}
R.~J. Scherrer and M.~S. Turner,
  \href{http://dx.doi.org/10.1103/PhysRevD.31.681}{``Decaying particles do not
  ``heat up'' the Universe,''{\em Phys. Rev. D} {\bfseries 31} (Feb, 1985)
  681--688}. \url{https://link.aps.org/doi/10.1103/PhysRevD.31.681}.

\bibitem{scherrer1988i}
R.~J. {Scherrer} and M.~S. {Turner},
  \href{http://dx.doi.org/10.1086/166534}{``{Primordial Nucleosynthesis with
  Decaying Particles. I. Entropy-producing Decays},''{\em The Astrophysical
  Journal} {\bfseries 331} (Aug., 1988)  19}.

\bibitem{scherrer1988ii}
R.~J. {Scherrer} and M.~S. {Turner},
  \href{http://dx.doi.org/10.1086/166535}{``{Primordial Nucleosynthesis with
  Decaying Particles. II. Inert Decays},''{\em The Astrophysical Journal}
  {\bfseries 331} (Aug., 1988)  33}.

\bibitem{audren2014}
B.~Audren, J.~Lesgourgues, G.~Mangano, P.~D. Serpico, and T.~Tram, ``{Strongest
  model-independent bound on the lifetime of Dark Matter},''
  \href{http://dx.doi.org/10.1088/1475-7516/2014/12/028}{{\em JCAP} {\bfseries
  12} (2014)  028}, \href{http://arxiv.org/abs/1407.2418}{{\ttfamily
  arXiv:1407.2418 [astro-ph.CO]}}.

\bibitem{poulin2016}
V.~Poulin, P.~D. Serpico, and J.~Lesgourgues, ``{A fresh look at linear
  cosmological constraints on a decaying dark matter component},''
  \href{http://dx.doi.org/10.1088/1475-7516/2016/08/036}{{\em JCAP} {\bfseries
  08} (2016)  036}, \href{http://arxiv.org/abs/1606.02073}{{\ttfamily
  arXiv:1606.02073 [astro-ph.CO]}}.

\bibitem{nygaard2020}
A.~Nygaard, T.~Tram, and S.~Hannestad, ``{Updated constraints on decaying cold
  dark matter},'' \href{http://dx.doi.org/10.1088/1475-7516/2021/05/017}{{\em
  JCAP} {\bfseries 05} (2021)  017},
  \href{http://arxiv.org/abs/2011.01632}{{\ttfamily arXiv:2011.01632
  [astro-ph.CO]}}.

\bibitem{Alvi:2022aam}
S.~Alvi, T.~Brinckmann, M.~Gerbino, M.~Lattanzi, and L.~Pagano, ``{Do you smell
  something decaying? Updated linear constraints on decaying dark matter
  scenarios},'' \href{http://arxiv.org/abs/2205.05636}{{\ttfamily
  arXiv:2205.05636 [astro-ph.CO]}}.

\bibitem{berezhiani2015}
Z.~Berezhiani, A.~D. Dolgov, and I.~I. Tkachev, ``{Reconciling Planck results
  with low redshift astronomical measurements},''
  \href{http://dx.doi.org/10.1103/PhysRevD.92.061303}{{\em Phys. Rev. D}
  {\bfseries 92} (2015) no.~6, 061303},
  \href{http://arxiv.org/abs/1505.03644}{{\ttfamily arXiv:1505.03644
  [astro-ph.CO]}}.

\bibitem{bringmann2018}
T.~Bringmann, F.~Kahlhoefer, K.~Schmidt-Hoberg, and P.~Walia, ``{Converting
  nonrelativistic dark matter to radiation},''
  \href{http://dx.doi.org/10.1103/PhysRevD.98.023543}{{\em Phys. Rev. D}
  {\bfseries 98} (2018) no.~2, 023543},
  \href{http://arxiv.org/abs/1803.03644}{{\ttfamily arXiv:1803.03644
  [astro-ph.CO]}}.

\bibitem{pandey2019}
K.~L. Pandey, T.~Karwal, and S.~Das, ``{Alleviating the $H_0$ and $\sigma_8$
  anomalies with a decaying dark matter model},''
  \href{http://dx.doi.org/10.1088/1475-7516/2020/07/026}{{\em JCAP} {\bfseries
  07} (2020)  026}, \href{http://arxiv.org/abs/1902.10636}{{\ttfamily
  arXiv:1902.10636 [astro-ph.CO]}}.

\bibitem{chen2020}
{\bfseries DES} Collaboration, A.~Chen {\em et al.}, ``{Constraints on dark
  matter to dark radiation conversion in the late universe with DES-Y1 and
  external data},'' \href{http://arxiv.org/abs/2011.04606}{{\ttfamily
  arXiv:2011.04606 [astro-ph.CO]}}.

\bibitem{blackadder2014}
G.~Blackadder and S.~M. Koushiappas, ``{Dark matter with two- and many-body
  decays and supernovae type Ia},''
  \href{http://dx.doi.org/10.1103/PhysRevD.90.103527}{{\em Phys. Rev. D}
  {\bfseries 90} (2014) no.~10, 103527},
  \href{http://arxiv.org/abs/1410.0683}{{\ttfamily arXiv:1410.0683
  [astro-ph.CO]}}.

\bibitem{blackadder2016}
G.~Blackadder and S.~M. Koushiappas, ``{Cosmological constraints to dark matter
  with two- and many-body decays},''
  \href{http://dx.doi.org/10.1103/PhysRevD.93.023510}{{\em Phys. Rev. D}
  {\bfseries 93} (2016) no.~2, 023510},
  \href{http://arxiv.org/abs/1510.06026}{{\ttfamily arXiv:1510.06026
  [astro-ph.CO]}}.

\bibitem{vattis2019}
K.~Vattis, S.~M. Koushiappas, and A.~Loeb, ``{Dark matter decaying in the late
  Universe can relieve the H0 tension},''
  \href{http://dx.doi.org/10.1103/PhysRevD.99.121302}{{\em Phys. Rev. D}
  {\bfseries 99} (2019) no.~12, 121302},
  \href{http://arxiv.org/abs/1903.06220}{{\ttfamily arXiv:1903.06220
  [astro-ph.CO]}}.

\bibitem{clark2020}
S.~J. Clark, K.~Vattis, and S.~M. Koushiappas, ``{Cosmological constraints on
  late-Universe decaying dark matter as a solution to the $H_0$ tension},''
  \href{http://dx.doi.org/10.1103/PhysRevD.103.043014}{{\em Phys. Rev. D}
  {\bfseries 103} (2021) no.~4, 043014},
  \href{http://arxiv.org/abs/2006.03678}{{\ttfamily arXiv:2006.03678
  [astro-ph.CO]}}.

\bibitem{haridasu2020}
B.~S. Haridasu and M.~Viel, ``{Late-time decaying dark matter: constraints and
  implications for the $H_0$-tension},''
  \href{http://dx.doi.org/10.1093/mnras/staa1991}{{\em Mon. Not. Roy. Astron.
  Soc.} {\bfseries 497} (2020) no.~2, 1757--1764},
  \href{http://arxiv.org/abs/2004.07709}{{\ttfamily arXiv:2004.07709
  [astro-ph.CO]}}.

\bibitem{abellan2021i}
G.~F. Abell\'an, R.~Murgia, and V.~Poulin, ``{Linear cosmological constraints
  on 2-body decaying dark matter scenarios and robustness of the resolution to
  the $S_8$ tension},'' \href{http://arxiv.org/abs/2102.12498}{{\ttfamily
  arXiv:2102.12498 [astro-ph.CO]}}.

\bibitem{abellan2022}
G.~F. Abell\'an, R.~Murgia, V.~Poulin, and J.~Lavalle, ``{Implications of the
  $S_8$ tension for decaying dark matter with warm decay products},''
  \href{http://dx.doi.org/10.1103/PhysRevD.105.063525}{{\em Phys. Rev. D}
  {\bfseries 105} (2022) no.~6, 063525},
  \href{http://arxiv.org/abs/2008.09615}{{\ttfamily arXiv:2008.09615
  [astro-ph.CO]}}.

\bibitem{Simon:2022ftd}
T.~Simon, G.~Franco~Abell\'an, P.~Du, V.~Poulin, and Y.~Tsai, ``{Constraining
  decaying dark matter with BOSS data and the effective field theory of
  large-scale structures},'' \href{http://arxiv.org/abs/2203.07440}{{\ttfamily
  arXiv:2203.07440 [astro-ph.CO]}}.

\bibitem{Anchordoqui2020}
L.~A. Anchordoqui, ``{Decaying dark matter, the $H_0$ tension, and the lithium
  problem},'' \href{http://dx.doi.org/10.1103/PhysRevD.103.035025}{{\em Phys.
  Rev. D} {\bfseries 103} (2021) no.~3, 035025},
  \href{http://arxiv.org/abs/2010.09715}{{\ttfamily arXiv:2010.09715
  [hep-ph]}}.

\bibitem{clark2022}
S.~J. Clark, K.~Vattis, J.~Fan, and S.~M. Koushiappas, ``{The $H_0$ and $S_8$
  tensions necessitate early and late time changes to $\Lambda$CDM},''
  \href{http://arxiv.org/abs/2110.09562}{{\ttfamily arXiv:2110.09562
  [astro-ph.CO]}}.

\bibitem{hannestad1998i}
S.~Hannestad, ``{Probing neutrino decays with the cosmic microwave
  background},'' \href{http://dx.doi.org/10.1103/PhysRevD.59.125020}{{\em Phys.
  Rev. D} {\bfseries 59} (1999)  125020},
  \href{http://arxiv.org/abs/astro-ph/9903475}{{\ttfamily
  arXiv:astro-ph/9903475}}.

\bibitem{kaplinghat1999}
M.~Kaplinghat, R.~E. Lopez, S.~Dodelson, and R.~J. Scherrer, ``{Improved
  treatment of cosmic microwave background fluctuations induced by a late
  decaying massive neutrino},''
  \href{http://dx.doi.org/10.1103/PhysRevD.60.123508}{{\em Phys. Rev. D}
  {\bfseries 60} (1999)  123508},
  \href{http://arxiv.org/abs/astro-ph/9907388}{{\ttfamily
  arXiv:astro-ph/9907388}}.

\bibitem{abellan2021neutrino}
G.~F. Abell\'an, Z.~Chacko, A.~Dev, P.~Du, V.~Poulin, and Y.~Tsai, ``{Improved
  cosmological constraints on the neutrino mass and lifetime},''
  \href{http://arxiv.org/abs/2112.13862}{{\ttfamily arXiv:2112.13862
  [hep-ph]}}.

\bibitem{chen2022}
J.~Z. Chen, I.~M. Oldengott, G.~Pierobon, and Y.~Y.~Y. Wong, ``{Weaker yet
  again: mass spectrum-consistent cosmological constraints on the neutrino
  lifetime},'' \href{http://arxiv.org/abs/2203.09075}{{\ttfamily
  arXiv:2203.09075 [hep-ph]}}.

\bibitem{barenboim2020}
G.~Barenboim, J.~Z. Chen, S.~Hannestad, I.~M. Oldengott, T.~Tram, and Y.~Y.~Y.
  Wong, ``{Invisible neutrino decay in precision cosmology},''
  \href{http://dx.doi.org/10.1088/1475-7516/2021/03/087}{{\em JCAP} {\bfseries
  03} (2021)  087}, \href{http://arxiv.org/abs/2011.01502}{{\ttfamily
  arXiv:2011.01502 [astro-ph.CO]}}.

\bibitem{chacko2020}
Z.~Chacko, A.~Dev, P.~Du, V.~Poulin, and Y.~Tsai, ``{Cosmological Limits on the
  Neutrino Mass and Lifetime},''
  \href{http://dx.doi.org/10.1007/JHEP04(2020)020}{{\em JHEP} {\bfseries 04}
  (2020)  020}, \href{http://arxiv.org/abs/1909.05275}{{\ttfamily
  arXiv:1909.05275 [hep-ph]}}.

\bibitem{blinov2020}
N.~Blinov, C.~Keith, and D.~Hooper, ``Warm decaying dark matter and the hubble
  tension,'' \href{http://dx.doi.org/10.1088/1475-7516/2020/06/005}{{\em
  Journal of Cosmology and Astroparticle Physics} {\bfseries 2020} (2020)
  no.~06, 005–005}.

\bibitem{CLASS2}
D.~Blas, J.~Lesgourgues, and T.~Tram, ``{The Cosmic Linear Anisotropy Solving
  System (CLASS). Part II: Approximation schemes},'' {\em JCAP} {\bfseries
  2011} (2011) no.~7, 034, \href{http://arxiv.org/abs/1104.2933}{{\ttfamily
  arXiv:1104.2933 [astro-ph.CO]}}.

\bibitem{anc}
L.~A. Anchordoqui, ``{Decaying dark matter, the $H_0$ tension, and the lithium
  problem},'' \href{http://dx.doi.org/10.1103/PhysRevD.103.035025}{{\em Phys.
  Rev. D} {\bfseries 103} (2021) no.~3, 035025},
  \href{http://arxiv.org/abs/2010.09715}{{\ttfamily arXiv:2010.09715
  [hep-ph]}}.

\bibitem{oldengott2014}
I.~M. Oldengott, C.~Rampf, and Y.~Y.~Y. Wong, ``{Boltzmann hierarchy for
  interacting neutrinos I: formalism},''
  \href{http://dx.doi.org/10.1088/1475-7516/2015/04/016}{{\em JCAP} {\bfseries
  04} (2015)  016}, \href{http://arxiv.org/abs/1409.1577}{{\ttfamily
  arXiv:1409.1577 [astro-ph.CO]}}.

\bibitem{oldengott2017}
I.~M. Oldengott, T.~Tram, C.~Rampf, and Y.~Y.~Y. Wong, ``{Interacting neutrinos
  in cosmology: exact description and constraints},''
  \href{http://dx.doi.org/10.1088/1475-7516/2017/11/027}{{\em JCAP} {\bfseries
  11} (2017)  027}, \href{http://arxiv.org/abs/1706.02123}{{\ttfamily
  arXiv:1706.02123 [astro-ph.CO]}}.

\bibitem{kolb}
E.~W. Kolb and M.~S. Turner,
  \href{http://dx.doi.org/10.1201/9780429492860}{{\em {The Early Universe}}},
  vol.~69.
\newblock 1990.

\bibitem{Ertas:2021xeh}
F.~Ertas, F.~Kahlhoefer, and C.~Tasillo, ``{Turn up the volume: listening to
  phase transitions in hot dark sectors},''
  \href{http://dx.doi.org/10.1088/1475-7516/2022/02/014}{{\em JCAP} {\bfseries
  02} (2022) no.~02, 014}, \href{http://arxiv.org/abs/2109.06208}{{\ttfamily
  arXiv:2109.06208 [astro-ph.CO]}}.

\bibitem{Escudero:2019gvw}
M.~Escudero and S.~J. Witte, ``{A CMB search for the neutrino mass mechanism
  and its relation to the Hubble tension},''
  \href{http://dx.doi.org/10.1140/epjc/s10052-020-7854-5}{{\em Eur. Phys. J. C}
  {\bfseries 80} (2020) no.~4, 294},
  \href{http://arxiv.org/abs/1909.04044}{{\ttfamily arXiv:1909.04044
  [astro-ph.CO]}}.

\bibitem{Escudero:2021rfi}
M.~Escudero and S.~J. Witte, ``{The hubble tension as a hint of leptogenesis
  and neutrino mass generation},''
  \href{http://dx.doi.org/10.1140/epjc/s10052-021-09276-5}{{\em Eur. Phys. J.
  C} {\bfseries 81} (2021) no.~6, 515},
  \href{http://arxiv.org/abs/2103.03249}{{\ttfamily arXiv:2103.03249
  [hep-ph]}}.

\bibitem{escudero2020}
M.~Escudero, J.~Lopez-Pavon, N.~Rius, and S.~Sandner, ``{Relaxing Cosmological
  Neutrino Mass Bounds with Unstable Neutrinos},''
  \href{http://dx.doi.org/10.1007/JHEP12(2020)119}{{\em JHEP} {\bfseries 12}
  (2020)  119}, \href{http://arxiv.org/abs/2007.04994}{{\ttfamily
  arXiv:2007.04994 [hep-ph]}}.

\bibitem{escudero2019}
M.~Escudero and M.~Fairbairn, ``{Cosmological Constraints on Invisible Neutrino
  Decays Revisited},''
  \href{http://dx.doi.org/10.1103/PhysRevD.100.103531}{{\em Phys. Rev. D}
  {\bfseries 100} (2019) no.~10, 103531},
  \href{http://arxiv.org/abs/1907.05425}{{\ttfamily arXiv:1907.05425
  [hep-ph]}}.

\bibitem{dodelson_widrow}
S.~Dodelson and L.~M. Widrow, ``{Sterile-neutrinos as dark matter},''
  \href{http://dx.doi.org/10.1103/PhysRevLett.72.17}{{\em Phys. Rev. Lett.}
  {\bfseries 72} (1994)  17--20},
  \href{http://arxiv.org/abs/hep-ph/9303287}{{\ttfamily arXiv:hep-ph/9303287}}.

\bibitem{hannestad2012}
S.~Hannestad, I.~Tamborra, and T.~Tram, ``{Thermalisation of light sterile
  neutrinos in the early universe},''
  \href{http://dx.doi.org/10.1088/1475-7516/2012/07/025}{{\em JCAP} {\bfseries
  07} (2012)  025}, \href{http://arxiv.org/abs/1204.5861}{{\ttfamily
  arXiv:1204.5861 [astro-ph.CO]}}.

\bibitem{white_paper_kev}
M.~Drewes {\em et al.}, ``{A White Paper on keV Sterile Neutrino Dark
  Matter},'' \href{http://dx.doi.org/10.1088/1475-7516/2017/01/025}{{\em JCAP}
  {\bfseries 01} (2017)  025},
  \href{http://arxiv.org/abs/1602.04816}{{\ttfamily arXiv:1602.04816
  [hep-ph]}}.

\bibitem{hannestad1998iii}
S.~Hannestad, ``{Constraining neutrino decays with CMBR data},''
  \href{http://dx.doi.org/10.1016/S0370-2693(98)00504-8}{{\em Phys. Lett. B}
  {\bfseries 431} (1998)  363--367},
  \href{http://arxiv.org/abs/astro-ph/9804075}{{\ttfamily
  arXiv:astro-ph/9804075}}.

\bibitem{ma1995}
C.-P. Ma and E.~Bertschinger, ``{Cosmological perturbation theory in the
  synchronous and conformal Newtonian gauges},''
  \href{http://dx.doi.org/10.1086/176550}{{\em Astrophys. J.} {\bfseries 455}
  (1995)  7--25}, \href{http://arxiv.org/abs/astro-ph/9506072}{{\ttfamily
  arXiv:astro-ph/9506072}}.

\bibitem{gradshteyn}
I.~S. Gradshteyn and I.~M. Ryzhik, {\em Table of integrals, series, and
  products.}
\newblock Academic Press, Amsterdam, Netherlands, 7~ed., 2007.

\bibitem{abramowitz}
M.~{Abramowitz} and I.~A. {Stegun}, {\em Handbook of Mathematical Functions
  with Formulas, Graphs, and Mathematical Tables}.
\newblock Dover, New York City, ninth dover printing, tenth gpo printing~ed.,
  1964.

\bibitem{Press2007}
W.~H. Press, S.~A. Teukolsky, W.~T. Vetterling, and B.~P. Flannery, {\em
  Numerical Recipes 3rd Edition: The Art of Scientific Computing}.
\newblock Cambridge University Press, 3~ed., 2007.

\bibitem{CLASS4}
J.~Lesgourgues and T.~Tram, ``{The Cosmic Linear Anisotropy Solving System
  (CLASS) IV: efficient implementation of non-cold relics},'' {\em JCAP}
  {\bfseries 2011} (2011) no.~9, 032,
  \href{http://arxiv.org/abs/1104.2935}{{\ttfamily arXiv:1104.2935
  [astro-ph.CO]}}.

\bibitem{hou2011}
Z.~Hou, R.~Keisler, Lloyd, M.~Millea, and C.~Reichardt,
  \href{http://dx.doi.org/10.1103/physrevd.87.083008}{``How massless neutrinos
  affect the cosmic microwave background damping tail,''{\em Physical Review D}
  {\bfseries 87} (apr, 2013)  }.
  \url{https://doi.org/10.1103%2Fphysrevd.87.083008}.

\bibitem{boss2016}
{\bfseries BOSS} Collaboration, S.~Alam {\em et al.}, ``{The clustering of
  galaxies in the completed SDSS-III Baryon Oscillation Spectroscopic Survey:
  cosmological analysis of the DR12 galaxy sample},''
  \href{http://dx.doi.org/10.1093/mnras/stx721}{{\em Mon. Not. Roy. Astron.
  Soc.} {\bfseries 470} (2017) no.~3, 2617--2652},
  \href{http://arxiv.org/abs/1607.03155}{{\ttfamily arXiv:1607.03155
  [astro-ph.CO]}}.

\bibitem{beutler2011}
F.~Beutler, C.~Blake, M.~Colless, D.~H. Jones, L.~Staveley-Smith, L.~Campbell,
  Q.~Parker, W.~Saunders, and F.~Watson,
  \href{http://dx.doi.org/10.1111/j.1365-2966.2011.19250.x}{``The 6dF Galaxy
  Survey: baryon acoustic oscillations and the local Hubble constant,''{\em
  Monthly Notices of the Royal Astronomical Society} {\bfseries 416} (jul,
  2011)  3017--3032}. \url{https://doi.org/10.1111%2Fj.1365-2966.2011.19250.x}.

\bibitem{ross2014}
A.~J. Ross, L.~Samushia, C.~Howlett, W.~J. Percival, A.~Burden, and M.~Manera,
  ``{The clustering of the SDSS DR7 main Galaxy sample \textendash{} I. A 4 per
  cent distance measure at $z = 0.15$},''
  \href{http://dx.doi.org/10.1093/mnras/stv154}{{\em Mon. Not. Roy. Astron.
  Soc.} {\bfseries 449} (2015) no.~1, 835--847},
  \href{http://arxiv.org/abs/1409.3242}{{\ttfamily arXiv:1409.3242
  [astro-ph.CO]}}.

\bibitem{scolnic2017}
{\bfseries Pan-STARRS1} Collaboration, D.~M. Scolnic {\em et al.}, ``{The
  Complete Light-curve Sample of Spectroscopically Confirmed SNe Ia from
  Pan-STARRS1 and Cosmological Constraints from the Combined Pantheon
  Sample},'' \href{http://dx.doi.org/10.3847/1538-4357/aab9bb}{{\em Astrophys.
  J.} {\bfseries 859} (2018) no.~2, 101},
  \href{http://arxiv.org/abs/1710.00845}{{\ttfamily arXiv:1710.00845
  [astro-ph.CO]}}.

\bibitem{benevento2020}
G.~Benevento, W.~Hu, and M.~Raveri, ``{Can Late Dark Energy Transitions Raise
  the Hubble constant?},''
  \href{http://dx.doi.org/10.1103/PhysRevD.101.103517}{{\em Phys. Rev. D}
  {\bfseries 101} (2020) no.~10, 103517},
  \href{http://arxiv.org/abs/2002.11707}{{\ttfamily arXiv:2002.11707
  [astro-ph.CO]}}.

\bibitem{camarena2021}
D.~Camarena and V.~Marra, ``{On the use of the local prior on the absolute
  magnitude of Type Ia supernovae in cosmological inference},''
  \href{http://dx.doi.org/10.1093/mnras/stab1200}{{\em Mon. Not. Roy. Astron.
  Soc.} {\bfseries 504} (2021)  5164--5171},
  \href{http://arxiv.org/abs/2101.08641}{{\ttfamily arXiv:2101.08641
  [astro-ph.CO]}}.

\bibitem{Audren:2012wb}
B.~Audren, J.~Lesgourgues, K.~Benabed, and S.~Prunet, ``{Conservative
  Constraints on Early Cosmology: an illustration of the Monte Python
  cosmological parameter inference code},''
  \href{http://dx.doi.org/10.1088/1475-7516/2013/02/001}{{\em JCAP} {\bfseries
  1302} (2013)  001},
\href{http://arxiv.org/abs/1210.7183}{{\ttfamily arXiv:1210.7183
  [astro-ph.CO]}}.
%%CITATION = ARXIV:1210.7183;%%.

\bibitem{montepython}
T.~Brinckmann and J.~Lesgourgues, ``{MontePython 3: boosted MCMC sampler and
  other features},'' \href{http://arxiv.org/abs/1804.07261}{{\ttfamily
  arXiv:1804.07261 [astro-ph.CO]}}.

\bibitem{Herold:2021ksg}
L.~Herold, E.~G.~M. Ferreira, and E.~Komatsu, ``{New Constraint on Early Dark
  Energy from Planck and BOSS Data Using the Profile Likelihood},''
  \href{http://dx.doi.org/10.3847/2041-8213/ac63a3}{{\em Astrophys. J. Lett.}
  {\bfseries 929} (2022) no.~1, L16},
  \href{http://arxiv.org/abs/2112.12140}{{\ttfamily arXiv:2112.12140
  [astro-ph.CO]}}.

\bibitem{Gomez-Valent:2022hkb}
A.~G\'omez-Valent, ``{A fast test to assess the impact of marginalization in
  Monte Carlo analyses, and its application to cosmology},''
  \href{http://arxiv.org/abs/2203.16285}{{\ttfamily arXiv:2203.16285
  [astro-ph.CO]}}.

\bibitem{Hamann_2012}
J.~Hamann, \href{http://dx.doi.org/10.1088/1475-7516/2012/03/021}{``Evidence
  for extra radiation? Profile likelihood versus Bayesian posterior,''{\em
  Journal of Cosmology and Astroparticle Physics} {\bfseries 2012} (mar, 2012)
  021--021}. \url{https://doi.org/10.1088%2F1475-7516%2F2012%2F03%2F021}.

\bibitem{raveri2018}
M.~Raveri and W.~Hu, ``{Concordance and Discordance in Cosmology},''
  \href{http://dx.doi.org/10.1103/PhysRevD.99.043506}{{\em Phys. Rev. D}
  {\bfseries 99} (2019) no.~4, 043506},
  \href{http://arxiv.org/abs/1806.04649}{{\ttfamily arXiv:1806.04649
  [astro-ph.CO]}}.

\bibitem{akaike1974}
H.~Akaike, ``A new look at the statistical model identification,''
  \href{http://dx.doi.org/10.1109/TAC.1974.1100705}{{\em IEEE Transactions on
  Automatic Control} {\bfseries 19} (1974) no.~6, 716--723}.

\bibitem{Hannestad:2000wx}
S.~Hannestad, ``{Stochastic optimization methods for extracting cosmological
  parameters from cosmic microwave background radiation power spectra},''
  \href{http://dx.doi.org/10.1103/PhysRevD.61.023002}{{\em Phys. Rev. D}
  {\bfseries 61} (2000)  023002},
  \href{http://arxiv.org/abs/astro-ph/9911330}{{\ttfamily
  arXiv:astro-ph/9911330}}.

\bibitem{DES}
{\bfseries DES} Collaboration, T.~M.~C. Abbott {\em et al.}, ``{Dark Energy
  Survey Year 3 results: Cosmological constraints from galaxy clustering and
  weak lensing},'' \href{http://dx.doi.org/10.1103/PhysRevD.105.023520}{{\em
  Phys. Rev. D} {\bfseries 105} (2022) no.~2, 023520},
  \href{http://arxiv.org/abs/2105.13549}{{\ttfamily arXiv:2105.13549
  [astro-ph.CO]}}.

\bibitem{dlmf}
``NIST Digital Library of Mathematical Functions.'' Release 1.1.1 of
  2021-03-15.
\newblock \url{http://dlmf.nist.gov/}. F.~W.~J. Olver, A.~B. {Olde Daalhuis},
  D.~W. Lozier, B.~I. Schneider, R.~F. Boisvert, C.~W. Clark, B.~R. Miller,
  B.~V. Saunders, H.~S. Cohl, and M.~A. McClain, eds.

\bibitem{arfken}
G.~B. Arfken, H.~J. Weber, and F.~E. Harris, {\em Mathematical Methods for
  Physicists}.
\newblock Academic Press, 7~ed., 2013.

\end{thebibliography}\endgroup

\appendix
\section{Derivation of background equations of motion} \label{apA}
In this appendix, we present a derivation of the background equations (\ref{daughter_back}) and (\ref{daughter_dens}) of the combined dark radiation species, starting from the equations for the individual species in reference~\cite{barenboim2020}. With the latter, after discarding inverse decay and quantum statistics terms and taking the massless limit, we have
\begin{align} \label{background_eq_dr}
	\pdv{\overline{f}_{\text{dr}} (q_2)}{\tau} = \pdv{\overline{f}_{l} (q_2)}{\tau}+ \frac{1}{2} \pdv{\overline{f}_{\phi} (q_2)}{\tau} = \frac{\mathfrak{g}^2 a^2 m_{H}^2}{2 \pi q_2^2} \int_{ q_{1-}}^{\infty} \d q_1 \frac{q_1}{\epsilon_1} \overline{f}_H (q_1),
\end{align}
with the integral limit $q_{1-} = \left| a^2 m_{H}^2/4 q_2 - q_2 \right|$. Here, we employed the definition $\overline{f}_{\text{dr}} = (2\overline{f}_l + \overline{f}_\phi)/2$. From the above, one obtains (\ref{daughter_back}) with the usual definition $\Gamma = \mathfrak{g}^2/4\pi$.

To obtain the equation for the energy density, we integrate the above over $4\pi a^{-4} \d q_2 \ q_2^3$. The left hand side becomes the usual $\d \rho_{\text{dr}}/\d \tau + 4aH \rho_{\text{dr}}$, while the right hand side becomes
\begin{align*}
	4\pi a^{-4} \int_{0}^{\infty} \d q_2 q_2^3 \frac{\mathfrak{g}^2 a^2 m_{H}^2}{2\pi q_2^2} \int_{q_{1-}^{l}}^{\infty} \d q_1 \frac{q_1}{\epsilon_1} f (q_1) = \frac{2\mathfrak{g}^2 m_{H}^2 }{a^2} \int_{0}^{\infty} \d q_2  q_2  \int_{q_{1-}^{l}}^{\infty} \d q_1 \frac{q_1}{\epsilon_1} f(q_1).
\end{align*}
Now, the $\d q_1$ integral does not reduce on its own, and the lower integral bound $q_{1-}^{l}$ contains a $q_2$-dependence which prevents us from evaluating the $\d q_2$ integral. However, we can relax the lower bound to $0$ by introducing a Heavyside step function in the integrand,
\begin{align*}
	\int_{q_{1-}}^{\infty} \d q_1 = \int_{0}^{\infty} \d q_1 \Theta (q_1 - q_{1-}) \Theta (q_{1+} - q_1),
\end{align*}
where the latter is trivial in the massless limit, $q_{1+} = \infty$, but important nonetheless. Indeed, reference~\cite{barenboim2020} derive the identity (A.24),
\begin{align*}
	\Theta (q_1 - q_{1-}) \Theta (q_{1+} - q_1) = \Theta (q_2 - q_{2-}^{H}) \Theta (q_{2+}^{H} - q_2)
\end{align*}
which allows us to translate the $\d q_1$ integral bounds into bounds on the $\d q_2$ integral given by $q_{2\pm}^{H} =(\epsilon_1 \pm q_1)/2$. Ultimately, we get the conversion
\begin{align} \label{bounds_conversion}
	\int_{0}^{\infty} \d q_2 \int_{ q_{1-}^{l}}^{\infty} \d q_1 = \int_{0}^{\infty} \d q_1 \int_{ q_{2-}^{H}}^{q_{2+}^{H}} \d q_2 ,
\end{align}
where we note that the order of integration must be reversed since the $\d q_2$ bounds now depend on $q_1$. Using this, the right hand side of the equation of motion becomes 
\begin{align}
	\frac{2\mathfrak{g}^2 m_{H}^2 }{a^2} \int_0^{\infty} \d q_1 \frac{q_1}{\epsilon_1} f(q_1)  \int_{q_{2-}^{H}}^{q_{2+}^{H}} \d q_2  q_2 .\label{12}
\end{align}
Now we can carry out the $\d q_2$ integral explicitly,
\begin{align*}
	\int_{q_{2-}^{H}}^{q_{2+}^{H}} \d q_2  q_2 = \frac{(q_{2+}^{H})^2 - (q_{2-}^{H})^2}{2} = \frac{\epsilon_1^2 + q_1^2 + 2 \epsilon_1 q_1 - \epsilon_1^2 - q_1^2 + 2 \epsilon_1 q_1}{2^3} = \frac{\epsilon_1 q_1}{2}.
\end{align*}
Substituting this in (\ref{12}) gives for the right hand side,
\begin{align*}
	\frac{2 \mathfrak{g}^2 m_{H}^2 }{a^2} \int_0^{\infty} \d q_1 \frac{q_1}{\epsilon_1} f(q_1)  \int_{q_{2-}^{H}}^{q_{2+}^{H}} \d q_2  q_2 &= \frac{\mathfrak{g}^2 m_{H}^2 }{a^2} \int_0^{\infty} \d q_1 q_1^2 f(q_1) \\
	&=\frac{\mathfrak{g}^2 m_{H}^2 }{a^2} \frac{a^3}{4\pi } n_H ,
\end{align*}
where $n_H$ again denotes the particle number density of the decaying particle. Equating the right and left hand sides of the equation of motion now finally gives
\begin{align}
	\dot{\rho}_{\text{dr}} + 4 a H \rho_{\text{dr}} &= \frac{\mathfrak{g}^2 m_{H}^2 a}{4\pi} n = a \Gamma m_{H} n_H. \label{background_density_daughter}
\end{align}
Comparing this with the evolution of the density of the decaying particle (\ref{mother_dens}), we see that the total comoving energy density is conserved in the decaying sector.

\section{Approximate solution to background equations} \label{apA2}
In this appendix, we expound on the analytical solution presented in section~\ref{sec:2.2.1}. We assume a power law Universe $a(t) = \kappa t^{2/3+3w}$ with equation of state parameter $w$ and $t$ denoting cosmic time. To first order in $q_1/m_H a$, the warm decaying species reduces to decaying cold dark matter, and in that case, the momentum dependence disappears, and one obtains the concrete solution
\begin{align} \nonumber
	f_H(q,t) = f_H (q, t_i) \exp \left( - \Gamma (t_i - t) \right)
\end{align}
where $t_i$ denotes some reference time. Here, the time evolution and momentum dependence decouple, so one can integrate directly over momentum to obtain the evolution of the integrated quantities
\begin{align} \nonumber
	n_H (t) = n_H (t_i) \exp \left( - \Gamma (t_i - t) \right), \quad \rho_H (t) = \rho_H (t_i) \exp \left( - \Gamma (t_i - t) \right).
\end{align}
The decay product energy density can be obtained by integrating over (\ref{daughter_dens}),
\begin{align} \nonumber
	\rho_{\text{dr}} (t) = \rho_{\text{dr}} (t_i) \left( \frac{a(t_i)}{a(t)} \right)^4 + \frac{\Gamma}{a(t)^4} \int_{t_i}^{t} \d t' \rho_H (t') a(t')^4.
\end{align}
For a power law Universe $a(t) = \kappa t^{2/3+3w}$ for some constant $\kappa$, we find
\begin{align} \label{dcdm_solution}
	\rho_{\text{dr}} (t) =  \rho_{\text{dr}} (t_i) \left( \frac{a(t_i)}{a(t)} \right)^4 + \rho_H (t_i) \frac{a(t_i)^3}{a(t)^4} \kappa \Gamma \exp(\Gamma t_i) \left( t_i^{\frac{5 + 3w}{3+3w}} E_{\frac{-2}{3+3w}} (\Gamma t_i) -  t^{\frac{5 + 3w}{3+3w}} E_{\frac{-2}{3+3w}} (\Gamma t)\right),
\end{align}
where $E_k (x)$ denotes the generalized exponential integral of variable order $k$. We note that for $w=1/3$, corresponding to a radiation dominated Universe, the order of the exponential integrals become $k=-1/2$ and a series of identities relate them to the error function through which one recovers the solution found in reference~\cite{nygaard2020}. Taking the limit $t/t_i \rightarrow \infty$, corresponding to the case where the entire population has decayed away, we can write the above in terms of the contribution to the radiation energy density today, expressed as an equivalent neutrino number,
\begin{align}
	\Delta N_{\text{eff}} \equiv N_{\text{eff}} \frac{\rho_{\text{dr}} (t) a(t)^4}{\rho_{\nu,0}} = N_{\text{eff}} \frac{\widetilde{\rho}_{H,0}}{\rho_{\nu,0}} \kappa \Gamma\left( \frac{5 + 3w}{3+3w} \right) \Gamma^{\frac{-2}{3+3w}},
\end{align}
where $\Gamma(x)$ denotes the Gamma function, $\widetilde{\rho}_{H,0} \equiv \rho_H(t_\text{ini}) a_\text{ini}^3$ is the density of the decaying species today \emph{if it were cold and stable}, and we assume no initial population of decay products. Here, one notes the characteristic scaling $\Delta N_{\text{eff}} \propto \Gamma^{\frac{-2}{3+3w}}$: Fast decays yield small final state densities and vice versa, since the decay products redshift faster than the parent particle.

The DCDM approximation holds after the species has become non-relativistic. Since we restrict ourselves to the area in parameter space where the species decays only after this, we can assume that only a negligible amount of decays take place prior to the non-relativistic transition. Hence, we can take the reference time $t_i$ to equal the non-relativistic transition time $t_{\text{nr}}$, as defined through the scale factor $a_{\text{nr}} \equiv a(t_{\text{nr}})\approx 3.15 T /m$, with boundary condition $\rho_{\text{dr}} (t_\text{nr}) = 0$. Using this, we can directly relate the initial densities to the final densities, yielding a useful starting point for the shooting algorithm of \textsc{class} which iteratively adjusts the two in order to obtain self-consistency~\cite{audren2014}. Firstly, the energy density of the decaying species at the non-relativistic transition is evaluated by assuming $\rho_H \propto a^{-4}$ redshifting prior to an instantaneous transition, giving $\rho_H(t_{\text{nr}}) = \rho_H (t_{\text{ini}}) a_{\text{ini}}^4/a_{\text{nr}}^4$. With this, and for simplicity assuming no initial population of decay products, (\ref{dcdm_solution}) leads directly to (\ref{shooting_guess}). From the latter, one sees that the dependence on the assumed dominant equation of state parameter $w$ manifests mainly through the generalized exponential integrals which define the "shape" of the decay. Thus, in estimating the final density parameter, this dependence has only a small impact, inasmuch as the majority of the decay is not ongoing today. In practice, we find the values of the final density parameters to be within a factor $5$ at relevant parameter values when assuming radiation and matter dominance, respectively. In the numerical implementation, we have used $w=1/3$, corresponding to a radiation dominated background, since then the order of the generalized exponential integrals is $-1/2$, and they can be rewritten in terms of the complementary error function. This can be seen by relating $E_{-1/2}(x)$ to $E_{1/2}(x)$ using the recurrence relation of the generalized exponential integrals (8.19.12 of~\cite{dlmf}), relating the latter to the upper incomplete Gamma function $\Gamma(k,x)$ with equation (8.19) of~\cite{dlmf} and finally relating the upper incomplete Gamma function to the complementary error function, for example with equation (13.93) of~\cite{arfken}. In the end, we find
\begin{align}
	E_{-1/2} (x) = \frac{\exp (-x)}{x} + \frac{1}{2 x^{3/2}}\Gamma\left( \frac{1}{2},x \right) = \frac{\exp (-x)}{x} + \frac{\sqrt{\pi}}{2 x^{3/2}} \text{erfc} (\sqrt{x}) \nonumber.
\end{align}
Since the complementary error function is implemented in most numerical libraries, this is the form we use. Lastly, one could expand the generalized exponential integrals in $\Gamma t_{\text{nr}} \ll 1$ and $\Gamma t_{0} \gg 1$, corresponding to non-relativistic decays occurring before today, since this is the scope of the current work. However, we have found that considerable precision is lost when using large decay constants or small masses such that the decay occurs close to the non-relativistic transition, so we have used the full solution instead. As stated in the main text, we find that (\ref{shooting_guess}) predicts the correct final density within a factor $5$ or so.

\section{Derivation of perturbation equations of motion} \label{apB}
In this appendix, we derive the expression (\ref{col_int}) for the momentum averaged decay product collision term. To first order, the perturbed combined distribution function is 
\begin{align}
	f_{\text{dr}} \equiv \overline{f}_{\text{dr}} (q, \tau) (1+ \Psi_{\text{dr}} (\bm{k}, \bm{q}, \tau) ) (\bm{k}, \bm{q}, \tau ) &= \left(\overline{f}_l (1 + \Psi_l) + \frac{1}{2}\overline{f}_\phi(1 + \Psi_\phi) \right) \nonumber \\
	&= \overline{f}_l + \frac{1}{2} \overline{f}_\phi + \overline{f}_l \Psi_l + \frac{1}{2} \overline{f}_\phi \Psi_\phi, \nonumber
\end{align}
in Fourier space, implicitly defining the combined perturbation 
\begin{align} \nonumber
	\Psi_{\text{dr}} \equiv  \frac{2 \overline{f}_l \Psi_l + \overline{f}_\phi \Psi_\phi}{2 \overline{f}_l + \overline{f}_\phi}.
\end{align}

In reference~\cite{barenboim2020}, the full Boltzmann equation for $\Psi_l$ and $\Psi_\phi$ are given. Here, we carry out the momentum averaging and write out the Boltzmann equation for the integrated hierarchy (\ref{massless_hierarchy}), using the definition (\ref{mom_avg}). Taking the time derivative of this definition gives four terms,
\begin{align}
	\dot{F}_{\text{dr},\ell} (q_2) &=  \dot{r}_{\text{dr}} \frac{\int q_2^2 \d q_2 \ q_2 \overline{f}_{\text{dr}}(q_2) \Psi_{\text{dr},\ell} (q_2) }{\int q_2^2 \d q_2 \ q_2 \overline{f}_{\text{dr}} (q_2)} \nonumber \\
	&- r_{\text{dr}} \frac{\int q_2^2 \d q_2 \ q_2 \overline{f}_{\text{dr}}(q_2) \Psi_{\text{dr},\ell} (q_2) }{\left(\int q_2^2 \d q_2 \ q_2 \overline{f}_{\text{dr}} (q_2)\right)^2} \int q_2^2 \d q_2 \ q_2 \dot{\overline{f}}_{\text{dr}} (q_2) \nonumber \\
	&+  r_{\text{dr}} \frac{\int q_2^2 \d q_2 \ q_2 \dot{\overline{f}}_{\text{dr}}(q_2) \Psi_{\text{dr},\ell} (q_2) }{\int q_2^2 \d q_2 \ q_2 \overline{f}_{\text{dr}} (q_2)} \label{total_mom_avg} \\
	&+  r_{\text{dr}} \frac{\int q_2^2 \d q_2 \ q_2 \overline{f}_{\text{dr}}(q_2) \dot{\Psi}_{\text{dr},\ell} (q_2) }{\int q_2^2 \d q_2 \ q_2 \overline{f}_{\text{dr}} (q_2)} \nonumber.
\end{align}
A thorough calculation will show that the first two terms cancel; indeed, this is the reason for including the $r_\text{dr}$ factor in the definition, as first done by reference~\cite{kaplinghat1999}. Next, we will see that a part of the fourth term cancels the third term. Firstly, we note that the collision term receives two contributions. The last term contains the total derivative of $\Psi_{\text{dr},\ell}$: With the exception of the collision term, this integral is a standard calculation (e.g.~\cite{ma1995}) with the result shown in equations (\ref{massless_hierarchy}). As such, we focus instead on the momentum average of the collision term.

Noting that 
\begin{align} \nonumber
	\dv{\Psi_{\text{dr},\ell}}{\tau} = \frac{2 \overline{f}_l \dot{\Psi}_{l,\ell} + \overline{f}_\phi \dot{\Psi}_{\phi,\ell} }{2 \overline{f}_l + \overline{f}_\phi},
\end{align}
and writing the collision terms for the species $l$ and $\phi$ from~\cite{barenboim2020} in the massless limit, the combined momentum dependent collision term is
\begin{align}
 \mathcal{C}^{(1)}_{\ell} [\Psi_\text{dr}(q_2)] &= \frac{2 \overline{f}_l (q_2) \mathcal{C}^{(1)}_{\ell} [\Psi_l (q_2)] + \overline{f}_\phi (q_2) \mathcal{C}^{(1)}_{\ell} [\Psi_\phi (q_2)]}{2 \overline{f}_l (q_2) + \overline{f}_\phi(q_2)} \nonumber \\
 &= \frac{2 a^2 m_H \Gamma}{q_2^2 \overline{f}_\text{dr}(q_2)} \int_{ q_{1-}}^\infty \d q_1 \frac{q_1}{\epsilon_1} \overline{f}_H (q_1) \left( \Psi_{H,\ell} (q_1) P_\ell (\cos \alpha^*) - \Psi_{\text{dr},\ell} (q_2) \right) \label{F_col_term}
\end{align}
where we recall the definitions $\epsilon_1 = (a^2 m_H^2 + q_1^2)^{1/2}$, $\Gamma = \mathfrak{g}^2 m_H/4\pi$ and 
\begin{align} \label{cosalpha}
	q_{1-} = \left| \frac{a^2 m_H^2}{4q_2} - q_2 \right|, \qquad \cos \alpha^* = \frac{\epsilon_1}{q_1} - \frac{a^2 m_H^2}{2 q_1 q_2}.
\end{align}
In the integrand of (\ref{F_col_term}), we see two terms. In the last one, $\Psi_{\text{dr},\ell}$ is evaluated at $q_2$, and not the integration variable $q_1$. The integral over $q_1$ in that term is thus proportional to the derivative of the background distribution, as seen from (\ref{daughter_back}). Due to the negative sign, this exactly cancels the third term in the total derivative of the momentum averaged perturbation (\ref{total_mom_avg}). We therefore find that the only addition to the time derivative of the momentum averaged perturbations (relative to its free variant~\cite{ma1995}) is the first term in the integrand of (\ref{F_col_term}),
\begin{align} \nonumber
	\dot{F}_{\text{dr},\ell} (q_2) &\supset \left( \dv{F_\text{dr}}{\tau} \right)^{(1)}_{C,\ell} \equiv \frac{4\pi r_{\text{dr}}}{\rho_{\text{dr}}} \int \d q_2 \ q_2^3 \overline{f}_{\text{dr}}(q_2) \left(\frac{2 a^2 m_H \Gamma}{q_2^2 \overline{f}_\text{dr}(q_2)} \int_{ q_{1-}}^\infty \d q_1 \frac{q_1}{\epsilon_1} \overline{f}_H (q_1) \Psi_{H,\ell} (q_1) P_\ell (\cos \alpha^*) \right) \\
	&= \frac{8\pi a^2 m_H \Gamma}{\rho_{\text{crit}}} \int_0^\infty \d q_2 \int_{ q_{1-}}^\infty \d q_1 \frac{q_1 q_2}{\epsilon_1} \overline{f}_H (q_1) \Psi_{H,\ell} (q_1) P_\ell (\cos \alpha^*) \equiv \frac{8\pi a^2 m_H \Gamma}{\rho_{\text{crit}}} \mathcal{I} \nonumber.
\end{align}
As was done in appendix~\ref{apA}, we can extend the $\d q_1$ lower bound to zero if we introduce a Heavyside step function in the integrand to enforce the correct bounds. This time, we will rewrite it in terms of the argument of the Legendre polynomial, $\cos \alpha^*$ with relation (A.24) from reference~\cite{barenboim2020},
\begin{align}
	\int_{ q_{1-}}^{\infty} \d q_1 = \int_0^{\infty} \d q_1 \Theta (1 - \cos^2 \alpha^*). \nonumber
\end{align}
Using this, the integral part of the perturbation becomes
\begin{align}
	\mathcal{I} = \int_{0}^{\infty} \d q_2 \int_0^\infty \d q_1 \Theta (1 - \cos^2 \alpha^*) \frac{q_1 q_2}{\epsilon_1} \overline{f}_H (q_1) \Psi_{H,\ell} (q_1) P_\ell (\cos \alpha^*) \label{I1}.
\end{align}
By substituting the $q_2$ variable for $u \equiv \cos \alpha^*$, given in equation (\ref{cosalpha}), this gives 
\begin{align} \nonumber
	\mathcal{I} = \int_0^{\infty} \d q_1 \int_{u_-}^{u_+} \d u \ \Theta (1- u^2) \frac{q_1^2 a^4 m_{H}^4}{4\epsilon_1^4} \overline{f}_H (q_1) \Psi_{H,\ell} (q_1)  \frac{P_\ell (u)}{\left(1 - \frac{q_1}{\epsilon_1} u\right)^3}
\end{align}
with the $u$-bounds
\begin{align}
	u_- = \lim_{q_2 \rightarrow 0} u(q_2) = -\infty, \qquad
	u_+ =  \lim_{q_2 \rightarrow \infty} u(q_2) = \frac{\epsilon_1}{q_1} \nonumber.
\end{align}
Since $u_- < -1$ and $u_+ > +1$, the $\d u$ integral bounds are completely controlled by the Heaviside step function, allowing the simplification to
\begin{align}
	\mathcal{I} &= \int_0^{\infty} \d q_1   q_1^2 \overline{f}_H (q_1) \Psi_{H,\ell} (q_1) \frac{a^4 m_{H}^4}{4\epsilon_1^4} \int_{-1}^{+1} \d u \ \frac{P_\ell (u)}{\left(1 - \frac{q_1}{\epsilon_1} u\right)^3} \nonumber \\
	&\equiv \int_0^{\infty} \d q_1   q_1^2 \overline{f}_H (q_1) \Psi_{H,\ell} (q_1) \frac{a^4 m_{H}^4}{2\epsilon_1^4} \frac{\mathcal{F}_\ell (q_1 / \epsilon_1)}{\left( 1 - q_1^2 / \epsilon_1^2 \right)^2},  \label{Ifinished}
\end{align}
where we have defined the scattering kernel
\begin{align} \label{scattering_kernel}
	\mathcal{F}_\ell (x) = \frac{\left(1 - x^2\right)^2}{2} \int_{-1}^{+1} \d u \frac{P_\ell (u)}{\left(1 - x u\right)^3}.
\end{align}
Finally, we can reduce the kinematical factor in equation (\ref{Ifinished}),
\begin{align*}
	\frac{a^4 m_{H}^4}{2\epsilon_1^4} \frac{1}{\left( 1 - q_1^2 / \epsilon_1^2 \right)^2} = \frac{a^4 m_{H}^4}{2 \left( \epsilon_1^2 - q_1^2 \right)^2} = \frac{1}{2},
\end{align*}
giving
\begin{align} \nonumber
	\mathcal{I} = \frac{1}{2} \int_0^{\infty} \d q_1   q_1^2 \overline{f}_H (q_1) \Psi_{H,\ell} (q_1) \mathcal{F}_\ell (q_1 / \epsilon_1).
\end{align}
Hence, we get the collision term,
\begin{align}
	 \left( \dv{F_\text{dr}}{\tau} \right)^{(1)}_{C,\ell} = \frac{8\pi a^2 m_H \Gamma}{\rho_{\text{crit}}} \left( \frac{1}{2} \int_0^{\infty} \d q_1   q_1^2 \overline{f}_H (q_1) \Psi_{H,\ell} (q_1) \mathcal{F}_\ell (q_1 / \epsilon_1) \right).
\end{align}
This can be rewritten with the equation of motion for the decay product energy density (\ref{daughter_dens}), yielding the expression
\begin{align}
	 \left( \dv{F_\text{dr}}{\tau} \right)^{(1)}_{C,\ell}  = \dot{r}_{\text{dr}} \frac{\int_0^{\infty} \d q  \ q^2 \overline{f}_H (q) \Psi_{H,\ell} (q) \mathcal{F}_\ell (q / \epsilon)}{\int_0^{\infty} \d q \ q^2 \overline{f}_H (q)},
\end{align}
using $\dot{r}_{\text{dr}} = \rho_{\text{crit}}^{-1} \ \d (\rho_{\text{dr}} a^4) / \d \tau =  r_{\text{dr}} \ a m_{H} \Gamma n_H / \rho_{\text{dr}}$. This is the final expression for the decay product perturbations, and matches the result found by reference~\cite{blinov2020}. Lastly, we emphasize two advantages of evolving the equations using this expression:
\begin{itemize}
	\item The collision term is independent of the dark radiation background distribution function $\overline{f}_{\text{dr}}$. Therefore, instead of tracking the distribution function itself, one needs only to evolve the energy density $\rho_\text{dr}$ according to equation (\ref{daughter_dens}).
	\item By computing only the momentum averaged perturbations, one escapes the need to evolve the momentum dependent perturbed distribution functions on a momentum grid, significantly reducing the required computation time.
\end{itemize}

\section{Fluid approximation} \label{apC}
The fluid approximation for non-cold species, originally introduced in~\cite{CLASS4}, is an approximation used to truncate the massive relic hierarchy at a given time for each mode, in order to reduce computation time. More precisely, upon switching on the fluid approximation, all information of the distribution function perturbations $\Psi$ of the DWDM particle is given up; instead, one evolves the first three integrated perturbations $\delta, \theta, \sigma$, approximating a truncation correction in the $\sigma$ equation~\cite{CLASS4}. By default, this is done for a given $k$ mode when $k\tau > 32$ in \textsc{class}. Importantly, the equations used for $\delta$ and $\theta$ are the continuity and Euler equations, respectively (hence the name of the approximation). Although a usual continuity equation holds for the entire decaying sector, the energy in the decaying species is of course not conserved due to the decay, and the fluid equations must be modified accordingly. The continuity equation for the entire decaying sector is
\begin{align} \label{continuity_together}
	\dot{\delta} = - \left(1+\frac{P}{\rho} \right) \left( \theta + \frac{\dot{h}}{2} \right) - 3 H \left( \frac{\delta P}{\delta \rho} - \frac{P}{\rho} \right) \delta ,
\end{align}
with all quantities representing sums over the DWDM and dark radiation contributions,
\begin{align} \nonumber
	\rho = \rho_{H} + \rho_{\text{dr}}, \quad p = p_{H} + p_{\text{dr}}, \quad \delta = \frac{\delta \rho_{H} + \delta \rho_{\text{dr}} }{\rho_{H} + \rho_{\text{dr}}}.
\end{align}
From this, we may find an expression for the continuity equation of the DWDM overdensity $\delta_{H} = \delta \rho_{H} / \rho_{H}$ alone. Isolating $\delta_{H}$ in equation (\ref{continuity_together}) and using the equation of motion for the $\ell = 0$ moment of the dark radiation perturbation $\dot{F}_{\text{dr}, 0}$, one arrives at
\begin{align}
	\dot{\delta}_{H} = - \left(1 + w\right)\left( \theta_{H} + \frac{\dot{h}}{2} \right)  -3 H \left( \frac{\delta p_{H}}{\delta \rho_{H}} - w \right) \delta_{H} - \left( \dv{F_{\text{dr}}}{\tau} \right)_{C,0}^{(1)}, \nonumber
\end{align}
where $w$ denotes the DWDM equation of state parameter and the collision term is given by equation (\ref{col_int}). As expected, the collision term takes into account the energy that is removed from the DWDM perturbations due to the decays. A similar equation holds for $\theta_{H}$, whose loss is determined by the $\ell = 1$ collision term, and likewise for all higher moments.

Unfortunately, even with the correct conservation equations, the fluid approximation cannot work for the DWDM species. In order to compute the dark radiation collision terms, one needs the DWDM distribution function perturbations $\Psi_{H,\ell} (q)$ which are not tracked in the fluid approximation. The simplest way out is to estimate the dark radiation collision terms by approximating the DWDM particle as a non-relativistic DCDM particle, in which case one can show that~\cite{blinov2020}
\begin{align}
	\left( \dv{F_{\text{dr}}}{\tau} \right)_{C,0}^{(1)} \rightarrow \dot{r}_{\text{dr}} \delta_{H}, \quad \left( \dv{F_{\text{dr}}}{\tau} \right)_{C,1}^{(1)} \rightarrow \frac{\dot{r}_{\text{dr}} \theta }{k}, \quad \left( \dv{F_{\text{dr}}}{\tau} \right)_{C,\ell}^{(1)} \approx 0 \quad (\ell \geq 3). \nonumber 
\end{align}
This should work reasonably well if the DWDM particle is heavy or very non-relativistic. In either case, since the fluid approximation is usually turned on when the majority of the DWDM population is depleted, the collision terms should play little role in the evolution of the perturbations, and the DCDM approximation should suffice. An interesting side effect of the DCDM approximation will be that $\delta_{H}$ may now converge completely toward $\delta_{\text{cdm}}$.
\begin{figure}[tb]
	\centering 
	\includegraphics[width=\textwidth]{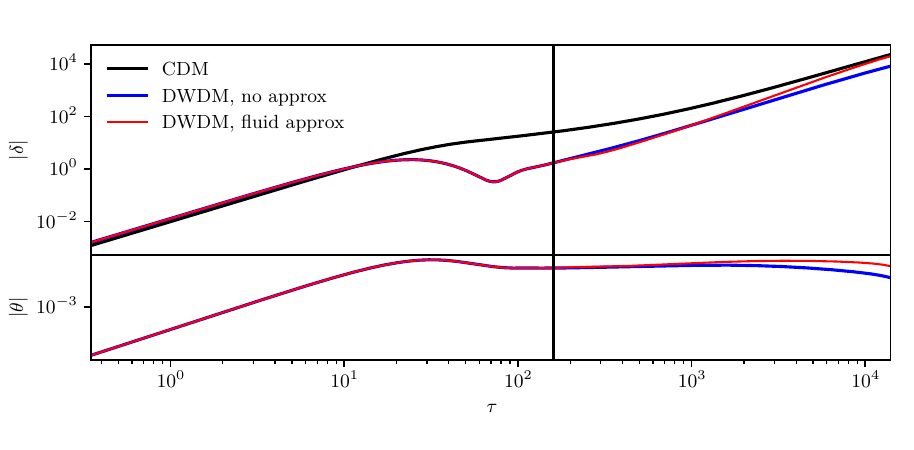}
	\caption{\label{perturbations_fluid} Same as figure~\ref{fig5}, but switching on the fluid approximation for the decaying species at $k\tau_0$, marked by the vertical black line. Due to the DCDM approximation, the decaying species' density perturbation converges strongly to that of cold dark matter, and the velocity divergence is diminished.}
\end{figure}

Figure~\ref{perturbations_fluid} shows the evolution of the density $\delta$ and velocity $\theta$ perturbations for the $k=0.2$ Mpc$^{-1}$ mode of a $m=10$ eV, $\Gamma= 10^8$ km s$^{-1}$ Mpc$^{-1}$ DWDM species as in figure~\ref{fig5}, but with the fluid approximation described in this section switched on at the default trigger value $k\tau_0 = 32$. Among other things, the DWDM overdensity is seen to converge to the DCDM faster than without the fluid approximation. In any case, applying the approximation will not invalidate any observable results, since the absolute perturbations $\delta \rho$, which are negligibly small sufficiently long after decay, generally enter the equations instead of the relative perturbations $\delta$. On the other hand, the increase in efficiency obtained from switching on the approximation is negligible; for the runs here, the difference in run time was less than $5 \%$ for a range of triggers. In conclusion, therefore, the fluid approximation, although applicable, was not found to be advantageous for the DWDM species.

\pagebreak
\section{Posterior distributions} \label{apD}
\begin{table}[h]
	\centering
	\begin{tabular}{? Sc | Sc | Sc | Sc ?}
		%\hline
		\specialrule{.12em}{0em}{0em} 
		\bf Parameter & DWDM & $\Delta N_\text{eff}$ & DCDM  \\
		\hline
		$10^{2}\omega_\text{b}$ & $2.246_{-0.021}^{+0.019}$ & $2.247_{-0.017}^{+0.015}$ & $2.247_{-0.016}^{+0.016}$ \\
		$\omega_\text{cdm}$ & $0.1216_{-0.0029}^{+0.0018}$ & $0.1215_{-0.0025}^{+0.0015}$ & $0.1209_{-0.0018}^{+0.0011}$ \\
		$H_0$ & $68.73_{-1.3}^{+0.81}$ & $68.66_{-1.0}^{+0.63}$ & $68.64_{-0.81}^{+0.45}$ \\
		$\ln 10^{10} A_s$ & $3.057_{-0.019}^{+0.016}$ & $3.056_{-0.016}^{+0.015}$ & $3.061_{-0.019}^{+0.012}$ \\
		$n_s$ & $0.9732_{-0.0081}^{+0.0059}$ & $0.9723_{-0.0061}^{+0.0049}$ & $0.9755_{-0.0072}^{+0.0042}$ \\
		$\tau_\text{reio}$ & $0.05779_{-0.0085}^{+0.0075}$ & $0.05764_{-0.0079}^{+0.0066}$ & $0.05971_{-0.0093}^{+0.0058}$ \\
		\hline
		$\sigma_8$ & $0.8186_{-0.021}^{+0.016}$ & $0.8174_{-0.0088}^{+0.0071}$ & $0.8224_{-0.011}^{+0.0061}$ \\
		\hline
		$\chi^2_\text{min} - \chi^2_{\text{min,}\Lambda\text{CDM}}$ & $0.22$ & $0.34$ & $0.68$ \\
		\specialrule{.12em}{0em}{0em} 
	\end{tabular}
	\caption{\label{tab:full} Results for base parameters from the MCMC runs for the DWDM, DCDM and $\Delta N_\text{eff}$ models using the baseline Planck 2018 + BAO dataset combination. The values represent mean values of the posteriors with 68 \% significance intervals.}
\end{table}

\begin{figure}[tb]
	\centering 
	\includegraphics[width=\textwidth]{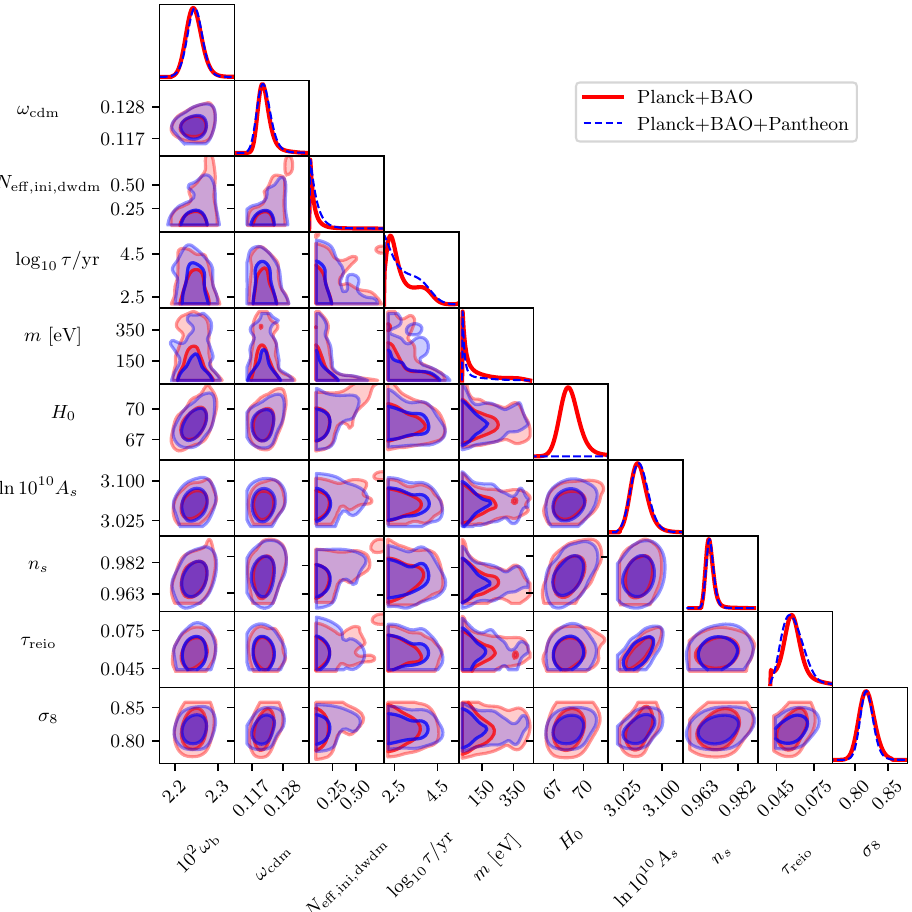}
	\caption{\label{mass_triangle} Triangle plot of the DWDM marginalized posteriors with Planck+BAO and Planck+BAO+Pantheon data.}
\end{figure}

\begin{figure}[tb]
	\centering 
	\includegraphics[width=\textwidth]{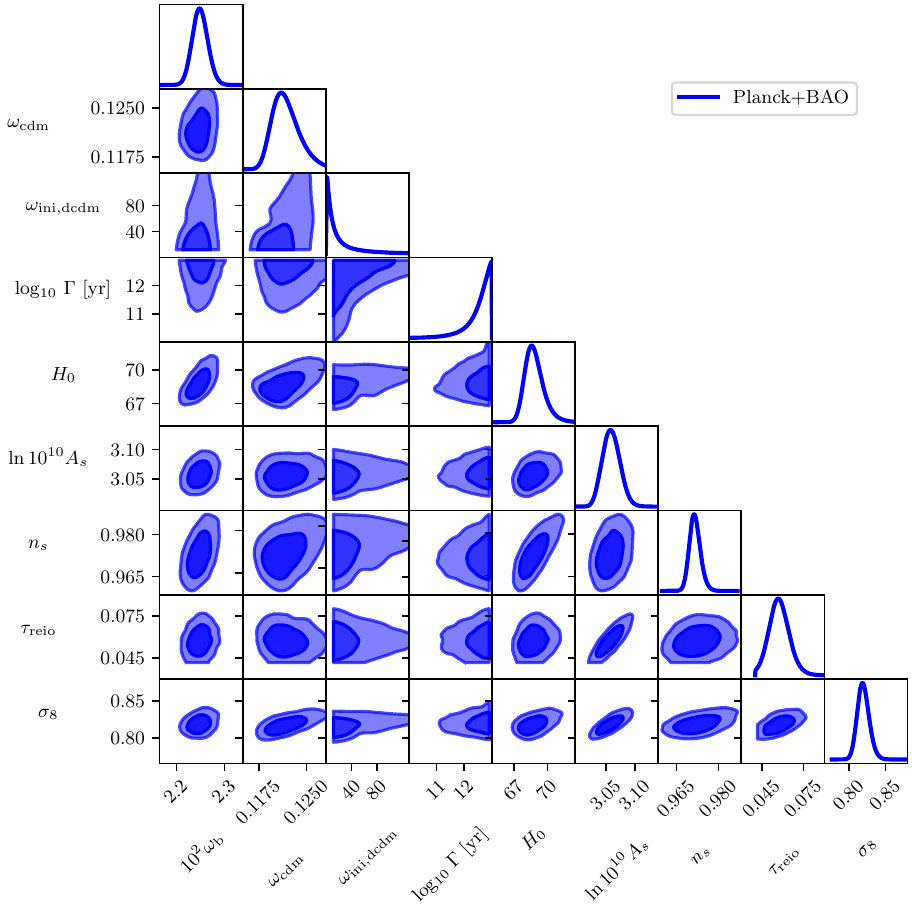}
	\caption{\label{dcdm_triangle} Triangle plot of the DCDM marginalized posteriors with Planck+BAO data. The prior bounds on $\log_{10} \Gamma$ has been chosen to match the prior used on the DWDM species and corresponds to a \textit{very} short-lived DCDM species. The abundance parameter $\omega_{\text{ini,dcdm}}$ is largely unconstrained since with these large decay constants, any decay radiation produced will redshift away before impacting any observable cosmology. We believe that the apparent preference for small $\omega_{\text{ini,dcdm}}$ values is a consequence of volume effects.}
\end{figure}

\begin{figure}[tb]
	\centering 
	\includegraphics[width=\textwidth]{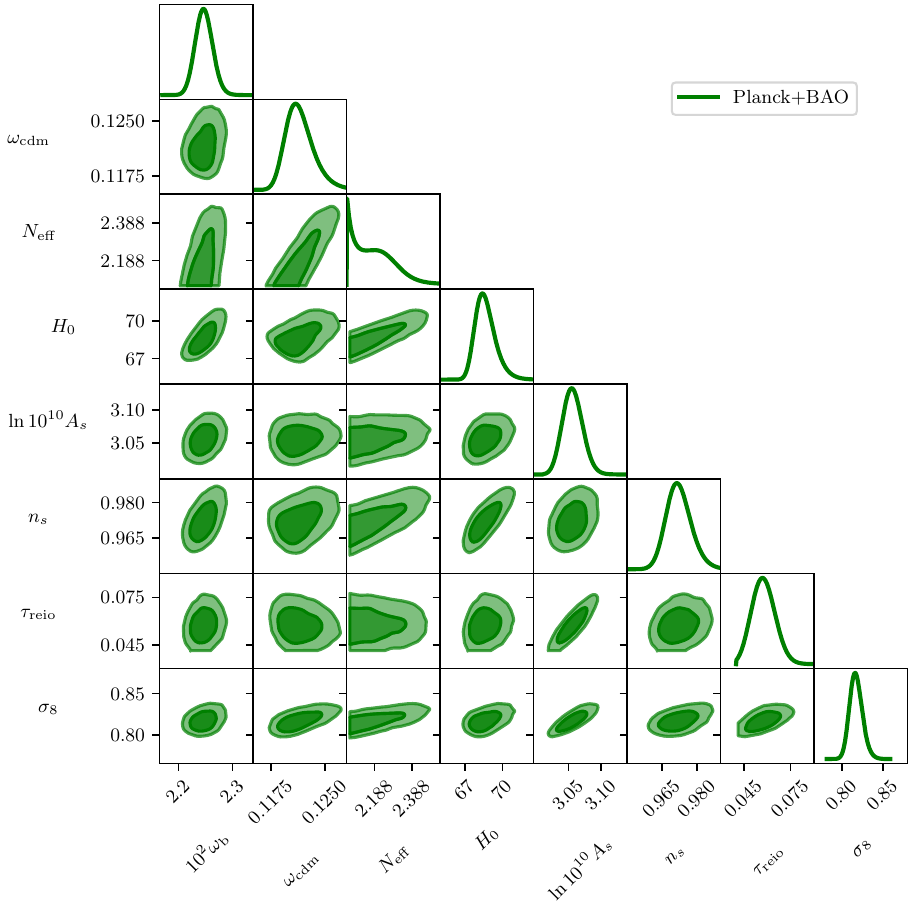}
	\caption{\label{neff_triangle} Triangle plot of the $\Delta N_\text{eff}$ marginalized posteriors with Planck+BAO data. Note that the $N_\text{eff}$ parameter has a lower bound of $2.038$, since we run \textsc{class} with one massive neutrino species and two massless ones, the latter contributing $N_\text{eff}=2.038$, such that the extra radiation is on top of the radiation density in the $\Lambda$CDM model.}
\end{figure}

%%%%%%%%%%%%
\end{document}